\documentclass[11pt,draftcls,onecolumn]{IEEEtran}

\usepackage{amsmath,amsfonts,amssymb,amsthm}
\usepackage{mathrsfs}
\usepackage{comment,blkarray}
\usepackage{multirow,bigdelim}
\usepackage{cite}
\usepackage[ruled,commentsnumbered, vlined]{algorithm2e}
\usepackage{tikz}
\usetikzlibrary{arrows,automata}
\usepackage[latin1]{inputenc}
\usepackage{verbatim}
\usepackage{graphicx}
\usepackage{subfigure}		
\usepackage{booktabs}
\usepackage{caption}													
\usepackage{enumitem}

\makeatletter
\newcommand{\rmnum}[1]{\romannumeral #1}
\newcommand{\Rmnum}[1]{\expandafter\@slowromancap\romannumeral #1@}

\newif\if@borderstar
\def\bordermatrix{\@ifnextchar*{%
  \@borderstartrue\@bordermatrix@i}{\@borderstarfalse\@bordermatrix@i*}%
}
\def\@bordermatrix@i*{\@ifnextchar[{\@bordermatrix@ii}{\@bordermatrix@ii[()]}}
\def\@bordermatrix@ii[#1]#2{%
\begingroup
  \m@th\@tempdima8.75\p@\setbox\z@\vbox{%
    \def\cr{\crcr\noalign{\kern 2\p@\global\let\cr\endline }}%
    \ialign {$##$\hfil\kern 2\p@\kern\@tempdima & \thinspace %
    \hfil $##$\hfil && \quad\hfil $##$\hfil\crcr\omit\strut %
    \hfil\crcr\noalign{\kern -\baselineskip}#2\crcr\omit %
    \strut\cr}}%
  \setbox\tw@\vbox{\unvcopy\z@\global\setbox\@ne\lastbox}%
  \setbox\tw@\hbox{\unhbox\@ne\unskip\global\setbox\@ne\lastbox}%
  \setbox\tw@\hbox{%
    $\kern\wd\@ne\kern -\@tempdima\left\@firstoftwo#1%
    \if@borderstar\kern2pt\else\kern -\wd\@ne\fi%
    \global\setbox\@ne\vbox{\box\@ne\if@borderstar\else\kern 2\p@\fi}%
    \vcenter{\if@borderstar\else\kern -\ht\@ne\fi%
    \unvbox\z@\kern -\if@borderstar2\fi\baselineskip}%
    \if@borderstar\kern -2\@tempdima\kern2\p@\else\,\fi\right\@secondoftwo#1 $%
  }\null \;\vbox{\kern\ht\@ne\box\tw@}%
\endgroup
}
\makeatother

\newtheorem{thm}{Theorem}

\newtheorem{lemma}[thm]{Lemma}
\newtheorem{eg}{Example}

\newtheorem{cor}[thm]{Corollary}

\newtheorem{defn}{Definition}
\newtheorem{rem}[thm]{Remark}

\allowdisplaybreaks

\newcommand{\w}{{\omega}}
\newcommand{\vb}{\vec{b}}

\newcommand{\vf}{\vec{f}}
\newcommand{\vk}{\vec{\ell}}

\newcommand{\vv}{\vec{v}}

\newcommand{\vh}{\vec{h}}
\newcommand{\vl}{\vec{k}}

\newcommand{\Fq}{\mathbb{F}_q}
\newcommand{\mB}{\mathcal{B}}
\newcommand{\mA}{\mathscr{A}}
\newcommand{\mE}{\mathscr{E}}
\newcommand{\mC}{\mathcal{C}}
\newcommand{\mL}{\mathcal{L}}
\newcommand{\mO}{\mathcal{O}}
\newcommand{\mK}{\mathcal{K}}

\newcommand{\bzero}{{\vec{0}}}

\newcommand{\Rank}{{\mathrm{Rank}}}
\newcommand{\tail}{{\mathrm{tail}}}
\newcommand{\head}{{\mathrm{head}}}
\newcommand{\Out}{{\mathrm{Out}}}
\newcommand{\In}{{\mathrm{In}}}

\newcommand{\CUT}{{\mathrm{CUT}}}
\newcommand{\mincut}{{\mathrm{mincut}}}

\begin{document}

\title{Local-Encoding-Preserving Secure Network Coding---Part \Rmnum{1}: Fixed Security Level
%\thanks{This paper was presented in part at the 2014 IEEE Information Theory Workshop.}
%This research is supported by the National Key Basic Research Program of China (973 Program Grant No. 2013CB834204), the National Natural Science Foundation of China (Nos. 61301137, 61171082) and the Fundamental Research Funds for Central Universities of China (No. 65121007). The materials in this paper were presented in part at both the IEEE Information Theory Workshop 2014, Tasmania, Australia, Nov. 2014, and the IEEE Information Theory Workshop 2015, Jeju Island, Korea, Oct. 2015.}
}

\author{Xuan~Guang,~\IEEEmembership{Member,~IEEE,}
        ~Raymond~W.~Yeung,~\IEEEmembership{Fellow,~IEEE,}
        ~and~Fang-Wei~Fu,~\IEEEmembership{Member,~IEEE}

%\thanks{ The problem of local-encoding-preserving secure network coding for a fixed security level and a flexible rate in the current paper was presented in part at 2014 IEEE Information Theory Workshop (ITW)~\cite{Guang-SNC-ITW14}, where the original results have been significantly improved in this paper.}
%\thanks{This work was partially supported by NSFC Grant (No.~61771259), the University Grants Committee of the Hong Kong SAR, China (Project No. AoE/E-02/08), and the Vice-Chancellor's One-off Discretionary Fund of CUHK (Project Nos. VCF2014030 and VCF2015007). This paper was presented in part at the 2014 IEEE Information Theory Workshop.}
    }

% The paper headers
\markboth{}%
{}
% The only time the second header will appear is for the odd numbered pages
% after the title page when using the twoside option.
%
% *** Note that you probably will NOT want to include the author's ***
% *** name in the headers of peer review papers.                   ***
% You can use \ifCLASSOPTIONpeerreview for conditional compilation here if
% you desire.

% If you want to put a publisher's ID mark on the page you can do it like
% this:
%\IEEEpubid{0000--0000/00\$00.00~\copyright~2007 IEEE}
% Remember, if you use this you must call \IEEEpubidadjcol in the second
% column for its text to clear the IEEEpubid mark.

% make the title area
\maketitle

\begin{abstract}
Information-theoretic security is considered in the paradigm of network coding in the presence of wiretappers, who can access one arbitrary edge subset up to a certain size, also referred to as the {\em security level}. Secure network coding is applied to prevent the leakage of the source information to the wiretappers. In this two-part paper, we consider the problem of secure network coding when the information rate and the security level can change over time.
In the current paper (i.e., Part~\Rmnum{1} of the two-part paper), we focus on the problem for a fixed security level and a flexible rate. To efficiently solve this problem, we put forward local-encoding-preserving secure network coding, where a family of secure linear network codes (SLNCs) is called {\em local-encoding-preserving} if all the SLNCs in this family share a common local encoding kernel at each intermediate node in the network. We present an efficient approach for constructing upon an SLNC that exists a local-encoding-preserving SLNC with the same security level and the rate reduced by one. By applying this approach repeatedly, we can obtain a family of local-encoding-preserving SLNCs with a fixed security level and multiple rates. We also develop a polynomial-time algorithm for efficient implementation of this approach. Furthermore, it is proved that the proposed approach incurs no penalty on the required field size for the existence of SLNCs in terms of the best known lower bound by Guang and Yeung. The result in this paper will be used as a building block for efficiently constructing a family of local-encoding-preserving SLNCs for all possible pairs of rate and security level, which will be discussed in the companion paper (i.e., Part~\Rmnum{2} of the two-part paper) \cite{part2}.
\end{abstract}
% Note that keywords are not normally used for peer review papers.
%\begin{IEEEkeywords}
%Secure network coding, information rate, security level, local-encoding-preserving property, algorithm
%\end{IEEEkeywords}

% For peer review papers, you can put extra information on the cover
% page as needed:
% \ifCLASSOPTIONpeerreview
% \begin{center} \bfseries EDICS Category: 3-BBND \end{center}
% \fi
%
% For peerreview papers, this IEEEtran command inserts a page break and
% creates the second title. It will be ignored for other modes.
\IEEEpeerreviewmaketitle

%%%%%%%%%%%%%%%%%%%%%%%%%%%%%%%%%-------Introduction------%%%%%%%%%%%%%%%%%
\section{Introduction}

In 1949, Shannon in his celebrated paper~\cite{Shannon-secrecy} put forward the well-known \textit{Shannon cipher system}. In this system, a sender wishes to transmit a private message to a receiver via a ``public'' channel which is eavesdropped by a wiretapper, and it is required that this wiretapper cannot obtain any information about the message. For this purpose, the sender applies a random key to encrypt the message and then transmit this encrypted message via the ``public'' channel. The random key is shared with the receiver via a ``secure'' channel that is inaccessible by the wiretapper. The receiver can recover the private message from the encrypted message and the random key, while the wiretapper cannot obtain any information about the private message. This is referred to as {\em information-theoretic security} in the literature.

In 1979, Blakley \cite{Blakley_secret-sharing-1979} and Shamir \cite{Shamir_secret-sharing-1979} independently put forward another well-known cipher system with information-theoretic security, called \textit{secret sharing}, which subsumes the Shannon cipher system. In this system, a secret is encoded into shares which are distributed among a set of participants, and it is required that only the qualified subsets of participants can recover the secret, while no information at all about the secret can be obtained from the shares of any unqualified set of participants.

Another related model called \textit{wiretap channel~\Rmnum{2}} was proposed by Ozarow and Wyner \cite{wiretap-channel-II}, in which the sender is required to transmit the message to the receiver through a set of noiseless point-to-point channels without leaking any information about the message to a wiretapper who can fully access any one but not more than one subset of the channels up to a certain size. Logically, wiretap channel \Rmnum{2} is a special case of secret sharing.

In 2000, Ahlswede~\textit{et al.}~\cite{Ahlswede-Cai-Li-Yeung-2000} formally put forward the concept of \textit{network coding} that allows the intermediate nodes in a noiseless network to process the received information. They proved that if coding is applied at the nodes, rather than routing only, the source node can multicast messages to each sink node at the theoretically maximum rate, i.e., the smallest minimum cut capacity between the source node and the sink nodes, as the alphabet size of both the information source symbol and the channel transmission symbol tends to infinity. Subsequently, Li~\textit{et~al.}~\cite{Li-Yeung-Cai-2003} proved that linear network coding with a finite alphabet is sufficient for optimal multicast by means of a vector space approach. Independently, Koetter and M\'{e}dard \cite{Koetter-Medard-algebraic} developed an algebraic characterization of linear network coding by means of a matrix approach. The above two approaches correspond to the {\em global} and {\em local} descriptions of linear network coding, respectively. Jaggi~\textit{et~al.} \cite{co-construction} further proposed a deterministic polynomial-time algorithm for constructing a linear network code. We refer the reader to \cite{Zhang-book, Yeung-book, Fragouli-book, Fragouli-book-app, Ho-book} for comprehensive discussions of network coding.

\subsection{Related Works}

In the paradigm of network coding, information-theoretic security is naturally considered in the presence of a wiretapper. This problem, called {\em secure network coding} problem, was introduced by Cai and Yeung in \cite{secure-conference, Cai-Yeung-SNC-IT}. In the model of secure network coding over a wiretap network, \rmnum{1}) the source node multicasts the source message to all the sink nodes which as legal users are required to decode the source message with zero error; and \rmnum{2}) the wiretapper, who can access any one wiretap set of edges, is not allowed to obtain any information about the source message. The foregoing three classical information-theoretically secure models~\cite{Shannon-secrecy, Blakley_secret-sharing-1979, Shamir_secret-sharing-1979, wiretap-channel-II} can be formulated as special cases of the above wiretap network model. In particular, a wiretap network is called a {\em $r$-wiretap network} if the wiretapper can fully access any one edge subset of size up to $r$, where $r$ is a nonnegative integer, called the {\em security level}.

Similar to the coding for the foregoing classical information-theoretically secure models \cite{Shannon-secrecy, Blakley_secret-sharing-1979, Shamir_secret-sharing-1979, wiretap-channel-II}, in secure network coding, it is necessary to randomize the source message to guarantee information-theoretic security. Cai and Yeung \cite{Cai-Yeung-SNC-IT} presented a code construction for $r$-wiretap networks. Subsequently, El~Rouayheb~\textit{et~al.}~\cite{Rouayheb-IT} showed that the Cai-Yeung  code construction can be viewed as a network generalization of the code construction for wiretap channel \Rmnum{2} in \cite{wiretap-channel-II}. Motivated by El~Rouayheb~\textit{et al.}, Silva and Kschischang \cite{Silva-UniversalSNC} proposed a universal design of secure network codes via rank-metric codes, in which the design of the linear network code for information transmission and the design of the code at the source node for information-theoretic security are over a finite field and its extension, respectively, so that the design of two codes can be separated. For secure network coding, the existing bounds on the required alphabet size in \cite{Cai-Yeung-SNC-IT, Rouayheb-IT, Silva-UniversalSNC} are roughly equal to the number of all wiretap sets, which is typically too large for implementation in terms of computational complexity and storage requirement. Recently, Guang and Yeung \cite{GY-SNC-Reduction} developed a systematic graph-theoretic approach to improve the required alphabet size for the existence of secure network codes and showed that the improvement in general is significant.

Secure network coding has also been investigated from different perspectives. Cheng and Yeung~\cite{Cheng-Yeung-Performance-SNC} studied the fundamental performance bounds for secure network coding in a wiretap network model where the collection of wiretap sets is arbitrary. Cui~\textit{et~al.}~\cite{Cui-Ho-Kliewer-SNC-Nonuniform} investigated the secure network coding problem in a single-source single-sink network with unequal channel capacities, where randomness is allowed to be generated at the non-source nodes. Furthermore, multi-source (multi-wiretapper) secure network coding problem was also investigated in the literature, such as \cite{Cai-Yeung-mul-source-SNC_ISIT07, Zhang-Yeung-mul-source-SNC_ISIT09, Chan-Grant-mul-Capacity-bound-SNC}.

For secure network coding, besides information-theoretic security discussed above, some other notions of security had also been considered in the literature. Bhattad and Narayanan \cite{Bhattad-Narayanan-Weakly-SNC_NetCod05} introduced {\em weakly secure network coding}, in which ``weak security'' is defined as the requirement that the wiretapper cannot recover any part of the source message. They also showed that we can use a weakly secure network code without trading off the rate, implying that a random key is not needed for weakly secure network coding. Another notion called {\em strong security} was introduced by Harada and Yamamoto~\cite{Harada-Yamamoto-Strongly-SNC_IEICE08}, where a network code is called \textit{$r$-strongly secure} if the wiretapper cannot obtain any information about any subset $\Delta$ of the $\w$ source symbols by accessing any edge subset of size up to  $\w+r-|\Delta|$. Subsequently, Cai~\cite{Cai-StronglyGenericLNC-ISIT09} proposed strongly generic linear network codes and proved that such codes are $r$-strongly secure.

Another line of research follows the so-called {\em Byzantine attacks}~\cite{Ho-Byzantine,Jaggi-Byzatine-ISIT07, Jaggi-Byzatine-IT08}, in which an adversary is able to modify the messages transmitted on the edges of a network. Another related model is the combination of secure network coding with network error correction coding, e.g., \cite{Ngai-Yeung-SNEC-NetCod09,Silva-UniversalSNC,Yao-Sliva-Jaggi-Lamgberg-SENC-ToN2014,
Zhuang-Luo-Vinck-Secure-NEC}, where the source message is required to be protected from both wiretapping and (random or malicious) errors. For an overview of secure network coding, we refer the reader to the two survey papers~\cite{Cai-Chan-Survey-SNC,Fragouli-Soljanin-DCC16}.

\subsection{Our Work}

In a secure network coding system, the requirements for information transmission and information security may vary. The information rate can change over time. For instance, the information source generated at the source node may have different rates at different times. The required security level can also change over time. For instance, the information source may be of different nature at different times, and so is the confidentiality associated with it. It is desirable to transmit a less confidential information source by using a secure network code with a lower security level, because a secure network code with a high security level requires more randomness for the key. Note that in a cipher system, randomness is a resource that needs to be optimized. Also as discussed above, there is an inherent tradeoff between the information rate and the security level of a secure network code. With all these considerations, the information rate and the security level of the system may need to be chosen differently at different times.

The straightforward approach is to use the existing code constructions to obtain a secure linear network code (SLNC) for each pair of rate and security level. However, this approach has a number of shortcomings. The construction of SLNCs for all the individual pairs of rate and security level incurs a high computation cost. Each node on the network needs to store the local encoding kernels for all the SLNCs, which incurs a high storage cost. The use of different SLNCs also causes an implementation overhead. To be specific, in using the SLNCs, the source node needs to inform each non-source node which rate and security level to use, and then each intermediate node needs to search for and apply the corresponding local encoding kernel for local encoding. This solution not only is cumbersome but also inefficient in terms of computation cost, storage cost, and implementation overhead.

To avoid the shortcomings of the above solution, in this paper we put forward {\em local-encoding-preserving secure network coding}, where a family of SLNCs is called {\em local-encoding-preserving} if all the SLNCs in this family share a common local encoding kernel at each intermediate node in the network. In other words, the same local encoding kernel is used at each intermediate node regardless of which SLNC in this family used. With this setup, the source node only needs to inform the sink nodes which SLNC in the family is in use, and there is no need to change the coding operations at the intermediate nodes.

%A property similar to local-encoding-preserving has been considered in  \cite{Fong-Yeung-variable-rate, Guang-uni-MDS, Jaggi-Byzatine-IT08} for different network coding problems. However, the main difficulty here is not to achieve the local-encoding-preserving property, which can be guaranteed easily (see Theorem~\ref{thm_LNC_transformation} in this paper). In fact, we can construct a family of local-encoding-preserving SLNCs by applying local-encoding-preserving property (guaranteed by Theorem~\ref{thm_LNC_transformation}) directly with the existing SLNC constructions. This approach, however, is inefficient and still has high computation cost and high storage cost at the source node. This will be discussed in detail later in this paper (i.e., Part~\Rmnum{1}) and in its companion (i.e., Part~\Rmnum{2})~\cite{part2}. In these two papers, we present systematic and efficient approaches to design families of local-encoding-preserving SLNCs for the different secure network coding systems discussed above.

Our idea is to build a new SLNC on an existing SLNC so that not only both SLNCs are local-encoding-preserving but also the new SLNC achieves another rate and security-level pair. To implement the local-encoding-preserving property, we consider two structures of SLNC constructions in the literature. One structure is to design an appropriate linear pre-coding operation at the source node upon a linear network code, where the linear pre-coding operation is designed for information security and the linear network code is constructed in advance for information transmission, e.g.,~\cite{Cai-Yeung-SNC-IT, Guang-opt-SNC}; and the other is to design an appropriate linear network code upon a linear pre-coding operation (regarded as a secure code) at the source node, e.g.,~\cite{Rouayheb-IT,Silva-UniversalSNC}. The later structure is seemly infeasible for implementing the local-encoding-preserving property because for this structure, once the secure code is changed for a different rate and/or a different security level, the linear network code which depends on the secure code has to be completely redesigned. In contrast, the former structure is feasible for implementing the local-encoding-preserving property. For this structure, since the linear pre-coding operation at the source node is built upon the linear network code constructed in advance, it is possible to only modify the linear pre-coding operation at the source node to adjust to the change of the rate and/or security level while preserving the local encoding kernels at the intermediate nodes. %In particular, it is easy to guarantee the local-encoding-preserving property when the dimension of the SLNCs is invariant.\footnote{This fact will be applied in solving the multiple rate and multiple security level secure network coding problem in Section~\ref{Sec_v-r_v-s-l}.} In general, when the dimensions of the SLNCs are variant, we need to consider the local-encoding-preserving property more.
Besides, it has been proved in~\cite{Cai-Yeung-SNC-IT, Guang-opt-SNC} that for this structure, the amount of randomness (entropy of the random key) required at the source node is the minimum possible for the required security level.

\begin{figure}[!t]
\centering
\begin{tikzpicture}[%Define standard arrow tip
>=stealth',
%Define style for different line styles
help lines/.style={dotted, thick},
axis/.style={<->},
important line/.style={thick},
connection/.style={thick, dashed}]
    % Axis
    \coordinate (y) at (0,6);
    \coordinate (x) at (6,0);
    \draw[<->] (y) node[above] {$r$} -- (0,0) node [below left]{$0$} --  (x) node[right]
    {$\w$};
    \path
    coordinate (start) at (0,5)
    coordinate (p00) at (0,0)
    coordinate (p01) at (0,1)
    coordinate (p02) at (0,2)
    coordinate (p03) at (0,3)
    coordinate (p04) at (0,4)
    coordinate (p10) at (1,0)
    coordinate (p11) at (1,1)
    coordinate (p12) at (1,2)
    coordinate (p13) at (1,3)
    coordinate (p14) at (1,4)
    coordinate (p20) at (2,0)
    coordinate (p21) at (2,1)
    coordinate (p22) at (2,2)
    coordinate (p23) at (2,3)
    coordinate (p30) at (3,0)
    coordinate (p31) at (3,1)
    coordinate (p32) at (3,2)
    coordinate (p40) at (4,0)
    coordinate (p41) at (4,1)
    coordinate (end) at (5,0);
    \draw[help lines] (start) node [left] {$C_{\min}$} -- (end) node[below] {$C_{\min}$};
    \draw[help lines] (p10) -- (p14);
    \draw[help lines] (p20) -- (p23);
    \draw[help lines] (p30) -- (p32);
    \draw[help lines] (p40) -- (p41);
    \draw[help lines] (p01) -- (p41);
    \draw[help lines] (p02) -- (p32);
    \draw[help lines] (p03) -- (p23);
    \draw[help lines] (p04) -- (p14);
    % Help coordinates for drawing the curve
     \filldraw [black]
     (start) circle (1pt)
     (p00) circle (1pt)
     (p01) circle (1pt)
     (p02) circle (1pt)
     (p03) circle (1pt)
     (p04) circle (1pt)
     (p10) circle (1pt)
     (p11) circle (1pt)
     (p12) circle (1pt)
     (p13) circle (1pt)
     (p14) circle (1pt)
     (p20) circle (1pt)
     (p21) circle (1pt)
     (p22) circle (1pt)
     (p23) circle (1pt)
     (p30) circle (1pt)
     (p31) circle (1pt)
     (p32) circle (1pt)
     (p40) circle (1pt)
     (p41) circle (1pt)
     (end) circle (1pt);
     %(slut) circle (1pt)
    %\filldraw [black]
\end{tikzpicture}
\caption{The rate and security-level region.}
\label{fig-region}
\end{figure}
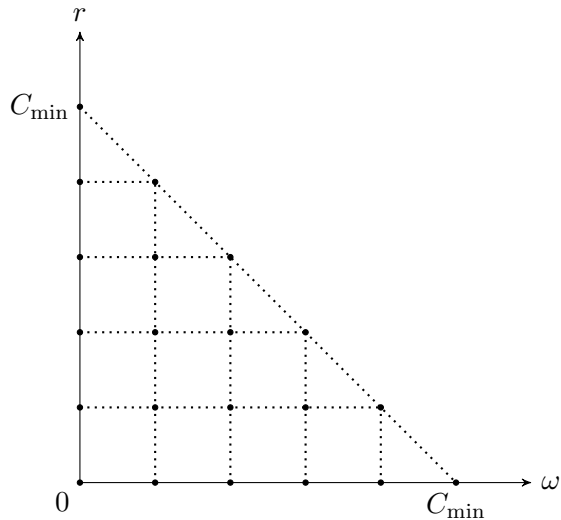

We divide the presentation of the results into two parts. In the current paper (i.e., Part~\Rmnum{1}), we design
\begin{enumerate}[label=(\roman*)]
  \item a family of local-encoding-preserving SLNCs for a fixed security level and a flexible rate.
\end{enumerate}
In the companion paper~\cite{part2} (i.e., Part~\Rmnum{2}), we design
\begin{enumerate}[label=(\roman*),resume]
  \item a family of local-encoding-preserving SLNCs for a fixed rate and a flexible security level;
  \item a family of local-encoding-preserving SLNCs for a fixed dimension (equal to the sum of rate and security level) and a flexible pair of rate and security level.
\end{enumerate}
It was proved in~\cite{Cai-Yeung-SNC-IT} that there exists an $n$-dimensional SLNC with rate $\w$ and security level $r$ (here $n=\w+r$) on the network $G$ if and only if $\w+r\leq C_{\min}$, where $C_{\min}$ is the smallest minimum cut capacity between the source node and each sink node. The set of all such rate and security-level pairs forms the \textit{rate and security-level region}, as depicted in Fig.~\ref{fig-region}. By combining the constructions of the 3 families of local-encoding-preserving SLNCs described above in suitable ways, we can design a family of local-encoding-preserving SLNCs that can be applied to all the pairs in the rate and security-level region. This will be discussed at the end of Part~\Rmnum{2} of the current paper~\cite{part2}.

%Guang~\textit{et al.}~\cite{Guang-SNC-ITW14, Guang-SNC-ITW15} considered designing local-encoding-preserving SLNCs for a fixed security level and multiple rates and for a fixed rate and multiple security levels, respectively, in which the results have already been significantly improved and covered in this paper and its companion~\cite{part2}.

\subsection{Organization and Contributions of this Paper}

The organization and main contributions of the paper are given as follows:
\begin{itemize}
  \item In Section~\ref{Sec_Preli}, we formally present the network model, linear network coding, and secure network coding with a construction of SLNC over a finite field of a reduced size. The necessary notation and definitions are also introduced.
  \item Section~\ref{Sec_v-r} is devoted to designing a family of local-encoding-preserving SLNCs for a fixed security level and a flexible rate. We present an efficient approach, which, upon any SLNC that exists, can construct a local-encoding-preserving SLNC with the same security level and the rate reduced by one. Then, starting with an SLNC with any fixed security level $r$ and the allowed maximum rate $C_{\min}-r$ and applying the proposed approach repeatedly, we can obtain a family of local-encoding-preserving SLNCs with the fixed security level $r$ and multiple rates from $C_{\min}-r$ to $0$.
  \item Although this approach gives the prescription for designing a local-encoding-preserving SLNC with the security level fixed and the rate reduced by one, it does not provide a method for efficient implementation. To tackle this problem, in Section~\ref{section_algorithm} we develop a polynomial-time algorithm for the efficient implementation of the approach. We also prove that the proposed approach and algorithm do not incur any penalty on the required field size for the existence of SLNCs in terms of the best known lower bound~\cite{GY-SNC-Reduction}.
  \item We conclude in Section~\ref{Sec_conclusion} with a summary of our results in the paper and an overview of the companion paper~\cite{part2}.
 \end{itemize}

\section{Preliminaries}\label{Sec_Preli}

\subsection{Network Model}

Let $G=(V, E)$ be a finite directed acyclic network with a single source $s$ and a set of sink nodes $T\subseteq V\setminus \{s\}$, where $V$ and $E$ are the sets of nodes and edges of $G$, respectively. For a directed edge~$e$ from node $u$ to node $v$, the node $u$ is called the {\em tail} of $e$ and the node $v$ is called the {\em head} of $e$, denoted by $\tail(e)$ and $\head(e)$, respectively. Further, for a node $u$, define $\In(u)$ as the set of incoming edges of $u$ and $\Out(u)$ as the set of outgoing edges of $u$. Formally, $\In(u)=\{e \in E:\ \head(e)=u\}$ and $\Out(u)=\{e\in E:\ \tail(e)=u\}$. Without loss of generality, assume that there are no incoming edges for the source node $s$ and no outgoing edges for any sink node $t\in T$. For convenience sake, however, we let $\In(s)$ be a set of $n$ {\em imaginary incoming edges}, denoted by $d_i$, $1\leq i \leq n$, terminating at the source node $s$ but without tail nodes, where the nonnegative integer $n$ is equal to the dimension of the network code in discussion. This will become clear later (see Definition~\ref{def-LNC}). Then, we see that $\In(s)=\big\{d_i:~1\leq i \leq n \big\}$. An index taken from an alphabet can be transmitted on each edge $e$ in $E$. In other words, the capacity of each edge is taken to be $1$. Parallel edges between two adjacent nodes are allowed.

In a network $G$, a {\em cut} between the source node $s$ and a non-source node $t$ is defined as a set of edges whose removal disconnects $s$ from $t$. The \textit{capacity} of a cut between $s$ and $t$ is defined as the number of edges in the cut, and the minimum of the capacities of all the cuts between $s$ and $t$ is called the \textit{minimum cut capacity} between them, denoted by $C_t$. A cut between $s$ and $t$ is called a \textit{minimum cut} if its capacity achieves the minimum cut capacity between them.

These concepts can be extended from a non-source node $t$ to an edge subset of $E$. We first need to consider a cut between~$s$ and a set of non-source nodes $T$ in the network $G$ as follows. We create a new node $t_T$, and for every node $t$ in $T$, add a new ``super-edge'' of infinite capacity\footnote{Infinite symbols in the alphabet can be transmitted by one use of the edge.} from $t$ to $t_T$ (which is equivalent to adding infinite parallel edges from $t$ to $t_T$). A cut of {\em finite} capacity between $s$ and $t_T$ is considered as a {\em cut} between $s$ and $T$. We can naturally extend the definitions of the capacity of a cut, the minimum cut capacity, and the minimum cut to the case of $T$. Now, we consider an edge subset $A\subseteq E$. For each edge $e\in A$, we introduce a node $t_e$ which splits $e$ into two edges $e^1$ and $e^2$ with $\tail(e^1)=\tail(e)$, $\head(e^2)=\head(e)$, and $\head(e^1)=\tail(e^2)=t_e$. Let $T_A=\{t_e: e\in A \}$. Then a {\em cut} between $s$ and $A$ is defined as a cut between $s$ and $T_A$, where, if $e^1$ or $e^2$ appears in the cut, replace it by~$e$. Similarly, the \textit{minimum cut capacity} between $s$ and $A$, denoted by $\mincut(s, A)$, is defined as the minimum cut capacity between $s$ and $T_A$, and a cut between $s$ and $A$ achieving the minimum cut capacity $\mincut(s, A)$ is called a \textit{minimum cut}.

\subsection{Linear Network Coding}

{\em Linear network coding} over a finite field is sufficient for achieving $C_{\min}\triangleq \min_{t\in T}C_t$, the theoretical maximum information rate for multicast \cite{Li-Yeung-Cai-2003, Koetter-Medard-algebraic}. We give the formal definition of a linear network code as follows.

\begin{defn}\label{def-LNC}
Let $\Fq$ be a finite field of order $q$, where $q$ is a prime power, and $n$ be a nonnegative integer. An $n$-dimensional $\Fq$-valued linear network code $\mC_n$ on the network $G=(V,E)$ consists of an $\Fq$-valued $\lvert \In(v)\rvert \times \lvert \Out(v)\rvert$ matrix $K_v=[k_{d,e}]_{d\in \In(v), e\in \Out(v)}$ for each non-sink node $v$ in $V$, i.e.,
\begin{align*}
\mC_n=\big\{ K_v:~v\in V\setminus T \big\},
\end{align*}
where $K_v$ is called the local encoding kernel of $\mC_n$ at $v$, and $k_{d,e}\in \Fq$ is called the local encoding coefficient for the adjacent edge pair $(d,e)$.
\end{defn}

For a linear network code $\mC_n$, the local encoding kernels $K_v$ at all the non-sink nodes $v\in V\setminus T$ induce an $n$-dimensional column vector $\vf^{\,(n)}_e$ for each edge $e$ in $E$, called the \textit{global encoding kernel} of~$e$, which can be calculated recursively according to a given ancestral order of edges in $E$ by
        \begin{align}\label{equ_ext_f}
        \vf^{\,(n)}_e=\sum_{d\in \In(\tail(e))}k_{d,e}\cdot\vf^{\,(n)}_d,
        \end{align}
with the boundary condition that $\vf_{d}^{\,(n)}$, $d\in \In(s)$ form the standard basis of the vector space $\Fq^n$. The set of global encoding kernels for all $e\in E$, i.e., $\big\{ \vf^{\,(n)}_e:~e\in E \big\}$, is also used to represent this linear network code $\mC_n$. However, we remark that a set of global encoding kernels $\big\{ \vf^{\,(n)}_e:~e\in E \big\}$ may correspond to more than one set of local encoding kernels $\big\{ K_v:~v\in V\setminus T \big\}$.

In using of this linear network code $\mC_n$, let $\mathbf{x}=\big(x_1 \ \ x_2 \ \ \cdots \ \ x_n \big) \in \Fq^{n}$ be the input of the source node~$s$. We assume that the input $\mathbf{x}$ is transmitted to $s$ through the $n$ imaginary incoming channels of the source node~$s$. Without loss of generality, $x_i$ is transmitted on the $i$th imaginary channel $d_i$, $1\leq i \leq n$. We use $y_e$ to denote the message transmitted on $e$, $\forall~e\in E$. Then $y_e$ can be calculated recursively by the equation
\begin{align}\label{equ_U_e}
y_e=\sum_{d\in \In(\tail(e))}k_{d,e}\cdot y_d
\end{align}
according to the given ancestral order of edges in $E$, with $y_{d_i}\triangleq x_{i}$, $1\leq i \leq n$. We see that $y_e$ in fact is a linear combination of the $n$ symbols $x_i$, $1\leq i\leq n$ of $\mathbf{x}$. It is readily seen that $y_{d_i}= \mathbf{x} \cdot \vf_{d_i}^{\,(n)}$ $(=x_{i})$, $1\leq i \leq n$. Then it can be shown by induction via \eqref{equ_ext_f} and \eqref{equ_U_e} that
\begin{align}
y_e= \mathbf{x} \cdot \vf_{e}^{\,(n)}, \quad \forall\,e\in E.
\end{align}
Furthermore, for each sink node $t\in T$, we define the matrix $F_t^{(n)}=\left[\vf_e^{\,(n)}:~e\in \In(t)\right]$. The sink node~$t$ can decode the source message with zero error if and only if $F_t^{(n)}$ is full rank, i.e., $\Rank\big(F_t^{(n)}\big)=n$. We say that an $n$-dimensional linear network code $\mC_n$ is {\em decodable} if for each sink node $t$ in $T$, the rank of the matrix $F_t^{(n)}$ is equal to the dimension $n$ of the code, i.e., $\Rank\big(F_t^{(n)}\big)=n$, $\forall~t\in T$. We refer the reader to \cite{Zhang-book, Yeung-book, Fragouli-book, Fragouli-book-app, Ho-book} for comprehensive discussions of linear network coding.

\subsection{Secure Network Coding}

Now, we present the secure network coding model. We assume that the source node $s$ generates a random source message $M$ taking values in the message set $\Fq^{\w}$ according to the uniform distribution, where the nonnegative integer~$\w$ is called the {\em information rate}. The source message $M$ needs to be multicast to each sink node $t\in T$, while being protected from a wiretapper who can access one but not more than one arbitrary edge subset of size not larger than $r$, where the nonnegative integer $r$ is called the {\em security level}. The network $G$ with a required security level $r$ is called an $r$-{\em wiretap network}. Similar to the other information-theoretically secure models, in our wiretap network model, it is necessary to randomize the source message to combat the wiretapper. The randomness available at the source node, called the \textit{key}, is a random variable $K$ that takes values in a set of keys $\Fq^r$ according to the uniform distribution.

We consider {\em secure linear network codes} (SLNCs) on an $r$-wiretap network $G$. Let $n=\w+r$, the sum of the information rate $\w$ and the security level $r$. An $\Fq$-valued $n$-dimensional SLNC on the $r$-wiretap network $G$ is
an $\Fq$-valued $n$-dimensional linear network on $G$ such that the following {\em decoding condition} and {\em security condition} are satisfied:
\begin{itemize}
  \item {\em decoding condition}: every sink node can decode the source message $M$ with zero error;
  \item {\em security condition}: the mutual information between $Y_A$ and $M$ is $0$, i.e., $I(Y_A ; M)=0$, for any edge subset $A\subseteq E$ with $\lvert A \rvert\leq r$, where we denote by $Y_e$ the random variable transmitted on the edge $e$ that is a linear function of the random source message $M$ and the random key $K$, and denote $(Y_e: e\in A)$ by $Y_A$ for an edge subset $A\subseteq E$.
  \end{itemize}
The nonnegative integers $\w$ and $r$ are also referred to as the {\em information rate} and {\em security level} of this $\Fq$-valued SLNC, respectively. The sum of the rate $\w$ and security level $r$ is the dimension $n$ of this $\Fq$-valued SLNC. In particular, it was proved in~\cite{Guang-opt-SNC} that $\Fq^r$ is the minimum set of keys to guarantee the security level of $r$ for any valid information rate $\w$ for $1\leq \w \leq C_{\min}-r$. When $r=0$, the secure network coding model reduces to the original network coding model.

\subsection{SLNC Construction}\label{subsec:SLNC_const}

We present the SLNC construction of Cai and Yeung~\cite{Cai-Yeung-SNC-IT}. Let $\w$ and $r$ be the information rate and the security level, respectively, and $n= \w+r\leq C_{\min}$. Let $\mC_n$ be an $n$-dimensional linear network code over a finite field $\Fq$ on the network $G$, of which all global encoding kernels are $\vf^{\,(n)}_e$, $e\in E$.

We use $\mE_r$ to denote the set of the edge subsets of size not larger than $r$, i.e., $\mE_r=\{ A\subseteq E:\ |A|\leq r \}$. Let $\vb_1^{\,(n)}$, $\vb_2^{\,(n)}$, $\cdots$, $\vb_\w^{\,(n)}$ be $\w$ linearly independent column vectors in $\Fq^n$ such that
\begin{equation}\label{secure_condition0}
\big\langle  \vb_i^{\,(n)}:\ 1\leq i \leq \w  \big\rangle \bigcap
\big\langle \vf^{\,(n)}_e:\ e\in A \big\rangle=\{\bzero\},\footnote{Here we use $\langle L \rangle$ to denote the subspace spanned by the vectors in a set $L$ of vectors. Furthemore, we always use $\bzero$ to denote an all-zero column vector in the paper, whose dimension is clear from the context.}\quad \forall~A\in \mE_r.
\end{equation}
Let $\vb_{\w+1}^{\,(n)}$, $\vb_{\w+2}^{\,(n)}$, $\cdots$, $\vb_n^{\,(n)}$ be another $n-\w$ column vectors in $\Fq^n$ such that $\vb_1^{\,(n)}$, $\vb_2^{\,(n)}$, $\cdots$, $\vb_n^{\,(n)}$ are linearly independent, and then let $Q^{(n)}=\left[ \vb_1^{\,(n)} \ \ \vb_2^{\,(n)} \ \  \cdots \ \ \vb_n^{\,(n)} \right]$.

Let $\mathbf{m}$, a row $\w$-vector in $\Fq^{\w}$, be the value of the source message $M$, and $\mathbf{k}$, a row $r$-vector in $\Fq^{r}$, be the value of the random key $K$. Then $\mathbf{x}=\big(\mathbf{m} \ \ \mathbf{k} \big)$ is the input of the source node $s$. At the source node $s$, $\mathbf{x}$ is first linearly encoded to $\mathbf{x}'=\mathbf{x}\cdot \big(Q^{(n)}\big)^{-1}$, and then apply the linear network code $\mC_n$ to multicast $\mathbf{x}'$ through the network $G$ to all the sink nodes. With this setting, it was proved in~\cite{Cai-Yeung-SNC-IT} that this coding scheme not only multicasts the source message $M$ to all the sink nodes at the rate $\w$ but also achieves the security level $r$. According to Theorem~\ref{thm_LNC_transformation} subsequently, this $n$-dimensional SLNC with the rate~$\w$ and security level $r$ is denoted by $\big(Q^{(n)}\big)^{-1} \cdot\,\mC_n$, and all the global encoding kernels of this SLNC are $\big(Q^{(n)}\big)^{-1} \cdot \vf^{\,(n)}_e$, $e\in E$.

\subsection{Field Size Reduction of the SLNC Construction}\label{subsec:SLNC_field_size}

Let $A\subseteq E$ be an edge subset. If $A$ satisfies $|A|=\mincut(s, A)$, then we say that $A$ is {\em regular}. Consider an arbitrary edge subset $A$ (not necessarily regular) with $\lvert A \rvert\leq r$, and replace $A$ by a minimum cut $\CUT_A$ between $s$ and $A$. Apparently, $\CUT_A$ is regular and still satisfies $\lvert \CUT_A \rvert \leq r$. A secure network code which is secure for $\CUT_A$ is also secure for $A$, namely that $I(Y_{\CUT_A} ; M)=0$ implies $I(Y_A ; M)=0$, since $Y_A$ is a deterministic function of $Y_{\CUT_A}$ by the mechanism of network coding. Thus, for the foregoing security condition, it suffices to consider all the regular edge subsets $A\subseteq E$ with $\lvert A \rvert \leq r$. Furthermore, for any regular edge subset $A\subseteq E$ with $\lvert A \rvert <r$, there must exist a regular edge subset $B$ with $\lvert B \rvert =r$ such that $A \subsetneq B$. By the same argument discussed above, it in fact suffices to consider all the regular edge subsets $A\subseteq E$ with $\lvert A \rvert =r$ for the security condition.

Let $A$ and $A'$ be two regular edge subsets with $\lvert A \rvert =\lvert A' \rvert =r$. Define a binary relation ``$\sim$'' between $A$ and $A'$: $A \sim A'$ if and only if there exists an edge set $\CUT$ which is a minimum cut between $s$ and $A$ and also between $s$ and $A'$, that is, $A$ and $A'$ have a common minimum cut between the source node $s$ and each of them. It was proved in Guang~\textit{et~al.}~\cite{Guang-SmlFieldSize-SNC-comm-lett} that ``$\sim$'' is an equivalence relation. With the relation ``$\sim$'', the collection of all the regular edge subsets of size $r$ can be partitioned into equivalence classes. Guang and Yeung~\cite{GY-SNC-Reduction} proved that each equivalence class contains a unique common minimum cut of all the edge subsets inside. Moreover, this common minimum cut is the {\em primary minimum cut} between~$s$ and any edge subset in this equivalence class, where a minimum cut between $s$ and an edge subset $A$ is {\em primary} if it separates $s$ and all the minimum cuts between $s$ and $A$. Such a primary minimum cut is unique and can be found in polynomial time.

Furthermore, we say an edge subset is {\em primary} if this edge subset is the primary minimum cut between~$s$ and itself. By using the foregoing argument again, we see that for the security condition, it suffices to consider all the primary edge subsets of size $r$. Precisely, we can replace \eqref{secure_condition0} by the following: we use $\mA_r$ to denote the set of the primary edge subsets of size $r$, i.e.,
\begin{align*}
\mA_r=\{ A\subseteq E:\ A \text{ is primary and } |A|=r \},\footnotemark
\end{align*}
\footnotetext{We refer the reader to Example~\ref{eg-1} in Section~\ref{section_algorithm} for illustrations of $\mA_r$.}
and let $\vb_1^{\,(n)}$, $\vb_2^{\,(n)}$, $\cdots$, $\vb_\w^{\,(n)}$ be $\w$ linearly independent column vectors in $\Fq^n$ such that
\begin{equation}\label{secure_condition}
\big\langle  \vb_i^{\,(n)}:\ 1\leq i \leq \w  \big\rangle \bigcap
\big\langle \vf^{\,(n)}_e:\ e\in A \big\rangle=\{\bzero\},\quad \forall~A\in \mA_r.
\end{equation}
It was also showed in~\cite{GY-SNC-Reduction} that $\lvert \mA_r \rvert$, as a new lower bound on the field size for the existence of SLNCs, improves the previous one ${|E| \choose r}$ (\,$\leq \lvert \mE_r \rvert$ clearly), and the improvement can be significant.

%%%%%%%%%%%%%%%%%%%%%%%%%%%%%%%%%%%%%%%%%%%%%%%%%%%%%%%%%%%%%%%%%%%%%%%%%%

\section{Secure Network Coding for Fixed Security Level and Flexible Rate}\label{Sec_v-r}

In this section, we consider the problem of designing a family of local-encoding-preserving SLNCs for a fixed information rate and a security level.

\subsection{Decoding Condition}

We first consider in this subsection the decoding condition under the preservation of the local encoding mappings. Let $n$ be a nonnegative integer not larger than $C_{\min}=\min_{t\in T} C_t$, and $\mC_n$ be an $n$-dimensional linear network code over a finite field $\Fq$ on the network $G$. First, we show the existence of a family of local-encoding-preserving decodable linear network codes in which the linear network codes have distinct dimensions.

\begin{thm}\label{thm_LNC_transformation}
Let $\mC_n=\big\{ K_v:~v\in V\setminus T \big\}$ be an $n$-dimensional decodable linear network code over a finite field $\Fq$ on the network $G=(V,E)$, of which the global encoding kernels are $\vf^{\,(n)}_e$, $e\in E$. Let $Q$ be an $m\times n$ ($m\leq n$) matrix over $\Fq$ and $Q\cdot\mC_n = \big\{ K_v^{(Q)}:~v\in V\setminus T \big\}$ with $K_s^{(Q)}=Q \cdot K_s$ and $K_v^{(Q)}=K_v$ for all $v\in V\setminus (\{s\}\cup T)$.
Then $Q \cdot \mC_n$ is an $m$-dimensional linear network code over $\Fq$ on $G$, of which the global encoding kernels are $Q \cdot \vf^{\,(n)}_e$, $e\in E$. This linear network code $Q\cdot\mC_n$ is called the transformation of $\mC_n$ by the matrix $Q$. In particular, $Q\cdot\mC_n$ is decodable provided that $Q$ is full row rank, i.e., $\Rank\big(Q\big)=m$.
\end{thm}
\begin{IEEEproof}
Clearly, $Q\cdot\mC_n$ is an $m$-dimensional linear network code over $\Fq$ on $G$ by Definition~\ref{def-LNC}. Let $\big\{ \vf^{\,(m)}_e(Q):~e\in E \big\}$ be the set of all the global encoding kernels with respect to the linear network code $Q\cdot\mC_n$. It suffices to prove that
\begin{align}\label{equ-thm_LNC_transformation}
\vf^{\,(m)}_e(Q)=Q\cdot\vf^{\,(n)}_e, \qquad \forall~e\in \Out(v)
\end{align}
for all non-sink nodes $v$, because for any $e\in E$, we have $e\in \Out(v)$ for some non-sink node $v$. This will be done by induction on the non-sink nodes according to any given ancestral order of the nodes in~$V$. First, consider the source node $s$ and obtain
\begin{align*}
\Big[\vf^{\,(m)}_e(Q):~e\in \Out(s)\Big]=K_s^{(Q)}=Q\cdot K_s=\Big[Q\cdot \vf^{\,(n)}_e:~e\in \Out(s)\Big].
\end{align*}
Now, consider an intermediate node $u$ in $V\setminus (\{s\}\cup T)$ and assume that \eqref{equ-thm_LNC_transformation} is satisfied for all non-sink nodes $v$ before $u$ according to the given ancestral order of the nodes. By the induction hypothesis, we have
\begin{align*}
&\Big[\vf^{\,(m)}_e(Q):~e\in \Out(u)\Big]=\Big[\vf^{\,(m)}_d(Q):~d\in \In(u)\Big]\cdot K_u^{(Q)}\\
&=\Big[Q \cdot \vf^{\,(n)}_d:~d\in \In(u)\Big]\cdot K_u=Q\cdot \Big[\vf^{\,(n)}_d:~d\in \In(u)\Big]\cdot K_u\\
&=Q \cdot \Big[\vf^{\,(n)}_e:~e\in \Out(u)\Big]=\Big[Q \cdot \vf^{\,(n)}_e:~e\in \Out(u)\Big].
\end{align*}
We thus proved that $\vf^{\,(m)}_e(Q)=Q\cdot\vf^{\,(n)}_e$ for all edges $e$ in $E$.

On the other hand, for each sink node $t$, the matrix $F_t^{(n)}$ in $\mC_{n}$ becomes the matrix
\begin{align*}
\left[Q\cdot\vf_e^{\,(n)}:~e\in \In(t) \right]=Q\cdot F_t^{(n)}
\end{align*}
in $Q\cdot \mC_n$. The decodability of $\mC_{n}$ implies that $\Rank\big(F_t^{(n)}\big)=n$. Hence, we see that $\Rank\big( Q\cdot F_t^{(n)} \big)=m$ provided that the matrix $Q$ is full row rank, i.e., $\Rank\big( Q \big)=m$. The theorem is proved.
\end{IEEEproof}

The following corollary is a straightforward application of Theorem~\ref{thm_LNC_transformation}.

\begin{cor}\label{thm_decoding_condition}
Let $\mC_n$ be an $n$-dimensional decodable linear network code  over the finite field $\Fq$ on the network $G=(V,E)$, of which the global encoding kernels are $\vf^{\,(n)}_e$, $e\in E$. Let $\vk$ be an arbitrary column $(n-1)$-vector in $\Fq^{n-1}$ and $I_{n-1}$ stand for the $(n-1)\times (n-1)$ identity matrix. Then \begin{align*}
\mC_{n-1}\triangleq\left[ I_{n-1} \ \ \vk\, \right]\cdot\mC_n
\end{align*}
is an $(n-1)$-dimensional decodable linear network code over $\Fq$ on $G$ with the same local encoding kernels at the intermediate nodes as the original code $\mC_n$, and all the global kernels of $\mC_{n-1}$ are
\begin{align}\label{equ_f_n-1}
\vf_e^{\,(n-1)}(\vk)\triangleq\left[ I_{n-1} \ \ \vk\, \right]\cdot\vf^{\,(n)}_e,\quad \forall~e\in E.
\end{align}
\end{cor}

\begin{rem}
Corollary~\ref{thm_decoding_condition} is essentially the same as Fong and Yeung \cite[Lemma~1]{Fong-Yeung-variable-rate}, in which this result was proved by using global encoding kernels. A similar but more complicated result in network error correction coding is Lemma~3 in \cite{Guang-uni-MDS}, which correspondingly was proved by using the extended global encoding kernels. %   have proved that $\mC_{n-1}=\big\{ \vf_e^{\,(n-1)}(\vk):~e\in E \big\}$ is an $(n-1)$-dimensional linear network code on $G$, and its local encoding coefficient of every adjacent edge pair $(d,e)$ can be preserved the   same as the original one of $(d,e)$ in $\mC_{n}$ (see the proof of \cite[Lemma~1]{Fong-Yeung-variable-rate}).
\end{rem}

%\begin{rem}
%It is worth mentioning that in use of the local-encoding-preserving $(n-1)$-dimensional linear network code $\mC_{n-1}$, we do not need to calculate all $\vf_e^{\,(n-1)}(\vk)$, $e\in E$. Instead, it suffices to calculate a new local encoding kernel at the source node $s$, i.e.,
%\begin{align}
%\Big[ I_{n-1} \ \ \vk \Big]\cdot K_s=\Big[ I_{n-1} \ \ \vk \Big]\cdot\Big[ %\vf^{\,(n)}_e:~e\in\Out(s)\Big]=\Big[\vf_e^{\,(n-1)}(\vk):~e\in \Out(s) \Big],
%\end{align}
%and then use the unchanged local encoding coefficients at each non-source node for encoding.
%\end{rem}

\subsection{Security Condition}\label{subsec_secure-cond}

In this subsection, we focus on how to guarantee the fixed security level for a family of local-encoding-preserving SLNCs with multiple rates.

Based on Corollary~\ref{thm_decoding_condition} and the construction of an SLNC in Sections~\ref{subsec:SLNC_const} and \ref{subsec:SLNC_field_size}, we can construct a family of local-encoding-preserving SLNCs with the same security level $r$ and information rates $\w$ from~$1$ to $C_{\min}-r$ as follows.\footnote{Under the requirement of the security level $r$, the allowed maximum information rate is $C_{\min}-r$.} Let $\mC_{C_{\min}}$ be a $C_{\min}$-dimensional decodable linear network code over $\Fq$ on $G$. Such a linear network code can be constructed in polynomial time (e.g.,~\cite{co-construction}). By Corollary~\ref{thm_decoding_condition}, we can obtain a family of local-encoding-preserving linear network codes $\big\{ \mC_n:~n=C_{\min}, C_{\min}-1, \cdots, r+1 \big\}$ of dimensions from $C_{\min}$ to $r+1$, all of which are decodable. Next, for each $n$-dimensional linear network code $\mC_n$ in this family, construct an $n\times n$ invertible matrix $Q^{(n)}$ satisfying \eqref{secure_condition}. Thus, we can obtain a family of SLNCs $\big\{ (Q^{(n)})^{-1}\cdot\,\mC_n:~n=C_{\min}, C_{\min}-1, \cdots, r+1 \big\}$ of the same security level $r$ and rates from $C_{\min}-r$ to $1$, and all of them have the same local encoding kernels at all the non-source nodes by Theorem~\ref{thm_LNC_transformation}. However, this approach not only requires the construction of the matrix $Q^{(n)}$ for each $n$, incurring a high computational complexity, but also requires the source node $s$ to store all the matrices $Q^{(n)}$ for each $n$.\footnote{\label{footnote}The computational complexity of the construction of $Q^{(n)}$ is shown to be $\mO\big(\w n^3|\mA_{r}|+\w n |\mA_{r}|^2+r n^2 \big)$ in Appendix~\ref{append-complx}, and the storage cost is $\mO\big( n^2 \big)$.}

To avoid these shortcomings, in the following we give a more efficient approach to solve the problem. We consider an $n$-dimensional linear network code $\mC_n$, of which all the global encoding kernels are $\vf^{\,(n)}_e$, $e\in E$. For any fixed column $(n-1)$-vector $\vk$, let
$$\mC_{n-1}=\left[ I_{n-1} \ \ \vk\, \right]\cdot\mC_n=\big\{ \vf_e^{\,(n-1)}(\vk):\ e\in E \big\}.$$
By Corollary~\ref{thm_decoding_condition}, $\mC_{n-1}$ is an $(n-1)$-dimensional linear network code and has the same local encoding kernels as $\mC_n$ at all the non-source nodes. On the other hand, by the construction of an SLNC in Sections~\ref{subsec:SLNC_const}~and~\ref{subsec:SLNC_field_size}, the $\w$ linearly independent column $n$-vectors $\vb_1^{\,(n)}$, $\vb_2^{\,(n)}$, $\cdots$, $\vb_\w^{\,(n)}$ satisfying~\eqref{secure_condition} guarantee the security level $r$. Our idea is to design an appropriate column $(n-1)$-vector $\vk$ to obtain $\w$ column $(n-1)$-vectors
\begin{align*}
\vb_i^{\,(n-1)}(\vk)=\left[ I_{n-1} \ \ \vk\, \right]\cdot \vb_i^{\,(n)},\quad 1\leq i \leq \w, \end{align*}
such that $(\w-1)$ vectors among them, e.g., $\vb_i^{\,(n-1)}(\vk)$, $1\leq i \leq \w-1$, are linearly independent, and the corresponding condition \eqref{secure_condition} for the rate $\w-1$ is satisfied, i.e.,    \begin{align}\label{secure_condition_2}
\big\langle  \vb_i^{\,(n-1)}(\vk):~1\leq i \leq \w-1 \big\rangle \bigcap \big\langle  \vf_e^{\,(n-1)}(\vk):~e\in A \big\rangle=\{\bzero\}, \quad \forall~A\in \mA_r.
\end{align}
With this, we can construct an $(n-1)$-dimensional SLNC that achieves a security level of $r$ and the lower information rate $\w-1$, and has the same local encoding kernels as $(Q^{(n)})^{-1}\cdot\mC_n$ at all the non-source nodes. Note that $K_s$ (of size $\w \times \lvert \Out(s)\rvert$), the local encoding kernel at the source node $s$, needs to be updated to $K_s^{(n-1)}(\vk)$ (of size $(\w-1) \times \lvert \Out(s)\rvert$). This will be discussed later.

Before presenting our approach, we first give some notations to be used frequently throughout the paper. Let $\mC_n$ be an $n$-dimensional decodable linear network code over a finite field $\Fq$ on the network~$G$, of which all the global encoding kernels are $\vf_e^{\,(n)}$, $e\in E$. Let $Q^{(n)}=\left[ \vb_1^{\,(n)} \ \  \vb_2^{\,(n)} \ \  \cdots \ \  \vb_n^{\,(n)} \right]$ be an $n\times n$ invertible matrix satisfying \eqref{secure_condition}, i.e.,
$(Q^{(n)})^{-1}\cdot\,\mC_n$ is an SLNC over $\Fq$ on $G$ with information rate~$\w$ and security level~$r$.

Define two types of vector spaces as follows:
\begin{align*}
\mL_A^{(n)}=&\big\langle  \vf_e^{\,(n)}:~e\in A \big\rangle, \qquad\  A\subseteq E\\
\mB_i^{(n)}=&\big\langle  \vb_j^{\,(n)}:~1\leq j \leq i \big\rangle, \quad 1\leq i \leq n.
\end{align*}
Furthermore, for a column $(n-1)$-vector $\vk\in \Fq^{n-1}$, recall that
\begin{align*}
\vf_e^{\,(n-1)}(\vk)&=\left[ I_{n-1} \ \ \vk\, \right]\cdot \vf_e^{\,(n)},\quad e\in E,\\
\vb_i^{\,(n-1)}(\vk)&=\left[ I_{n-1} \ \ \vk\, \right]\cdot \vb_i^{\,(n)},\quad 1\leq i \leq n,
\end{align*}
and define two types of vector spaces similarly:
\begin{align*}
\mL_A^{(n-1)}(\vk)&=\big\langle  \vf_e^{\,(n-1)}(\vk):~e\in A \big\rangle,\qquad\ A\subseteq E,\\
\mB_i^{(n-1)}(\vk)&=\big\langle  \vb_j^{\,(n-1)}(\vk):~1\leq j \leq i \big\rangle, \quad 1\leq i \leq n.
\end{align*}
In particular, for $\vk=\bzero$, let
\begin{align}
\vf_e^{\,(n-1)} & = \vf_e^{\,(n-1)}(\bzero)=\left[ I_{n-1} \ \ \bzero \right]\cdot \vf_e^{\,(n)},\qquad  e\in E,\label{vf-n-1}\\
\vb_i^{\,(n-1)} & = \vb_i^{\,(n-1)}(\bzero)=\left[ I_{n-1} \ \ \bzero \right]\cdot \vb_i^{\,(n)}, \qquad 1\leq i \leq n,\label{vb-n-1}
\end{align}
i.e., $\vf_e^{\,(n-1)}$ (resp. $\vb_i^{\,(n-1)}$) is the sub-vector of $\vf_e^{\,(n)}$ (resp. $\vb_i^{\,(n)}$) containing the first $(n-1)$ components, and
\begin{align}
\mL_A^{(n-1)}&=\big\langle  \vf_e^{\,(n-1)}:~e\in A \big\rangle,\qquad\  A\subseteq E,\label{L_A}\\
\mB_i^{(n-1)}&=\big\langle  \vb_j^{\,(n-1)}:~1\leq j \leq i \big\rangle, \quad 1\leq i \leq n \label{B_i}.
\end{align}

Now, we present our approach in detail. By \eqref{vb-n-1}, we first compute
\begin{align}\label{matrix_n-1}
\left[ \vb_1^{\,(n-1)} \ \ \vb_2^{\,(n-1)} \ \  \cdots \ \ \vb_\w^{\,(n-1)} \right]=
\left[ I_{n-1} \ \ \bzero \right]\cdot \left[ \vb_1^{\,(n)} \ \ \vb_2^{\,(n)} \ \  \cdots \ \ \vb_\w^{\,(n)} \right],
\end{align}
which, together with the linear independence of $\vb_1^{\,(n)}$, $\vb_2^{\,(n)}$, $\cdots$, $\vb_\w^{\,(n)}$, implies that
\begin{align*}
\Rank\Big( \left[ \vb_1^{\,(n-1)} \ \ \vb_2^{\,(n-1)} \ \  \cdots \ \ \vb_\w^{\,(n-1)} \right]\Big )\geq \w-1.
\end{align*}
In other words, there must exist $\w-1$ vectors out of $\vb_1^{\,(n)}$, $\vb_2^{\,(n)}$, $\cdots$, $\vb_\w^{\,(n)}$ are linearly independent. Without loss of generality, we assume that $\vb_1^{\,(n-1)}$, $\vb_2^{\,(n-1)}$, $\cdots$, $\vb_{\w-1}^{\,(n-1)}$ are linearly independent.

Furthermore, with the global encoding kernels $\vf_e^{\,(n)}$, $e\in E$ of $\mC_n$, we partition $\mA_r$ into the following two disjoint subsets:
\begin{align*}
\mA_r'  = \{A\in \mA_r:~\text{the $n-1$ vectors } \vb_i^{\,(n-1)}, 1\leq i \leq \w-1,\ \vf_e^{\,(n-1)}, e\in A, \text{ are linearly dependent}\},
\end{align*}
and
\begin{align*}
\mA_r''  =\{A\in \mA_r:~\text{the $n-1$ vectors } \vb_i^{\,(n-1)}, 1\leq i \leq \w-1,\ \vf_e^{\,(n-1)}, e\in A, \text{ are linearly independent}\}.
\end{align*}

The following two lemmas give the prescriptions of designing the column $(n-1)$-vectors $\vk$ for $\mA_r'$ and $\mA_r''$, respectively.

\begin{lemma}\label{lem_3}
Let $\vb_1^{\,(n-1)}$, $\vb_2^{\,(n-1)}$, $\cdots$, $\vb_{\w-1}^{\,(n-1)}$ be $\w-1$ linearly independent column $(n-1)$-vectors over a finite field $\Fq$, where $n=\w+r$. For any column $(n-1)$-vector $\vk\in \Fq^{n-1}$ such that
\begin{align}\label{vk_choosen_case1}
\vk\in \Fq^{n-1}\setminus \bigcup_{A\in \mA_r'} \Big( \mB_{\w-1}^{(n-1)}+\mL_A^{(n-1)} \Big),
\end{align}
then the following are satisfied:
\begin{itemize}
  \item the column $(n-1)$-vectors $\vb_i^{\,(n-1)}(\vk)$, $1\leq i \leq \w-1$, are linearly independent;
  \item $\mB_{\w-1}^{(n-1)}(\vk) \bigcap \mL_A^{(n-1)}(\vk)=\{ \bzero \}$, $\forall~A \in \mA_r'$.
\end{itemize}
\end{lemma}
\begin{IEEEproof}
Let $\vk$ be an arbitrary column $(n-1)$-vector satisfying \eqref{vk_choosen_case1}.
We first prove that $\vb_1^{\,(n-1)}(\vk)$, $\vb_2^{\,(n-1)}(\vk)$, $\cdots$, $\vb_{\w-1}^{\,(n-1)}(\vk)$ are linearly independent. Let $\vb_i^{\,(n)}=\big[ b_{i,1}\ b_{i,2}\ \cdots\ b_{i,n} \big]^\top$, $1\leq i \leq \w-1$. Assume that $\alpha_1$, $\alpha_2$, $\cdots$, $\alpha_{\w-1}$ are $\w-1$ elements in $\Fq$ such that
\begin{align}
\bzero=\sum_{i=1}^{\w-1}\alpha_i\vb_i^{\,(n-1)}(\vk)=\sum_{i=1}^{\w-1}\alpha_i(\vb_i^{\,(n-1)}+b_{i,n}\vk),
\end{align}
or equivalently,
\begin{align}\label{thm_secure_condition_case1_eq1}
\sum_{i=1}^{\w-1}\alpha_i\vb_i^{\,(n-1)}=-\Big(\sum_{i=1}^{\w-1}\alpha_ib_{i,n}\Big)\cdot\vk,
\end{align}
where we note that $b_{i,n}$ is the last component of $\vb_i^{\,(n)}$. By~\eqref{vk_choosen_case1}, we have $\vk\notin \mB_{\w-1}^{(n-1)}$, which, together with \eqref{thm_secure_condition_case1_eq1}, implies that $\sum_{i=1}^{\w-1}\alpha_i b_{i,n}=0$ and $\sum_{i=1}^{\w-1}\alpha_i\vb_i^{\,(n-1)}=\bzero$. Thus, we obtain
\begin{align*}
\sum_{i=1}^{\w-1}\alpha_i\vb_i^{\,(n)}=\sum_{i=1}^{\w-1}\alpha_i
\begin{bmatrix}\vb_i^{\,(n-1)}\\b_{i,n}\end{bmatrix}=\bzero,
\end{align*}
which implies that $\alpha_i=0$ for all $1\leq i \leq \w-1$ since $\vb_1^{\,(n)}$, $\vb_2^{\,(n)}$, $\cdots$, $\vb_{\w-1}^{\,(n)}$ are linearly independent. We thus have proved that $\vb_1^{\,(n-1)}(\vk)$, $\vb_2^{\,(n-1)}(\vk)$, $\cdots$, $\vb_{\w-1}^{\,(n-1)}(\vk)$ are linearly independent.

Next, we prove that $\mB_{\w-1}^{(n-1)}(\vk)\bigcap \mL_A^{(n-1)}(\vk)=\{\bzero\}$ for all $A\in \mA_r'$. We assume the contrary that there exists a wiretap set $A\in \mA_r'$ such that \begin{align}
\mB_{\w-1}^{(n-1)}(\vk)\bigcap\mL_A^{(n-1)}(\vk)\neq \{\bzero\}.
\end{align}
Let $\vv\in \Fq^{n-1}$ be a nonzero vector in this intersection.
Then, we have $\beta_1$, $\beta_2$, $\cdots$, $\beta_{\w-1}$ in $\Fq$, not all zero, such that
\begin{align}
\vv&=\sum_{i=1}^{\w-1}\beta_i\vb_i^{\,(n-1)}(\vk)
=\sum_{i=1}^{\w-1}\beta_i(\vb_i^{\,(n-1)}+b_{i,n}\vk),\label{equ1_thm1_case1}
\end{align}
and another $r$ elements in $\Fq$, denoted by $\gamma_e$, $e\in A$, which are not all zero, such that
\begin{align}
\vv&=\sum_{e\in A}\gamma_e\vf_e^{\,(n-1)}(\vk)
=\sum_{e\in A}\gamma_e(\vf_e^{\,(n-1)}+f_{e,n}\vk),\label{equ2_thm1_case1}
\end{align}
where $f_{e,n}$ is the last component of $\vf_e^{\,(n)}$. By \eqref{equ1_thm1_case1} and \eqref{equ2_thm1_case1}, we immediately obtain
\begin{align*}
\Bigg(\sum_{i=1}^{\w-1}\beta_i\vb_i^{\,(n-1)}-\sum_{e\in A}\gamma_e\vf_e^{\,(n-1)} \Bigg)+\Bigg(\sum_{i=1}^{\w-1}\beta_ib_{i,n}-\sum_{e\in A}\gamma_ef_{e,n} \Bigg)\cdot\vk=\bzero.
\end{align*}
Together with $\vk\notin \mB_{\w-1}^{(n-1)}+\mL_A^{(n-1)}$ from \eqref{vk_choosen_case1} (which implies $\vk\neq\bzero$), we further have
\begin{align}
\sum_{i=1}^{\w-1}\beta_ib_{i,n}-\sum_{e\in A}\gamma_ef_{e,n}=0 \qquad \text{ and }\qquad \sum_{i=1}^{\w-1}\beta_i\vb_i^{\,(n-1)}-\sum_{e\in A}\gamma_e\vf_e^{\,(n-1)}=\bzero,
\end{align}
or equivalently,
\begin{align}\label{eq18}
\sum_{i=1}^{\w-1}\beta_ib_{i,n}=\sum_{e\in A}\gamma_ef_{e,n} \qquad\text{ and }\qquad \sum_{i=1}^{\w-1}\beta_i\vb_i^{\,(n-1)}=\sum_{e\in A}\gamma_e\vf_e^{\,(n-1)}.
\end{align}
Now, we can write \eqref{eq18} in vector form as
\begin{align}
\sum_{i=1}^{\w-1}\beta_i\begin{bmatrix}\vb_i^{\,(n-1)}\\b_{i,n}\end{bmatrix}=\sum_{e\in A}\gamma_e\begin{bmatrix} \vf_e^{\,(n-1)}\\ f_{e,n}\end{bmatrix},
\end{align}
namely,
\begin{align}\label{3}
\sum_{i=1}^{\w-1}\beta_i\vb_i^{\,(n)}=\sum_{e\in A}\gamma_e\vf_e^{\,(n)}.
\end{align}
Together with the linear independence of $\vb_1^{\,(n)}$, $\vb_2^{\,(n)}$, $\cdots$, $\vb_{\w-1}^{\,(n)}$, we obtain
\begin{align}
\bzero\neq\sum_{i=1}^{\w-1}\beta_i\vb_i^{\,(n)}=\sum_{e\in A}\gamma_e\vf_e^{\,(n)},
\end{align}
since $\beta_1$, $\beta_2$, $\cdots$, $\beta_{\w-1}$ are not all zero. This implies that $\mB_{\w-1}^{(n)} \bigcap \mL_A^{(n)}\neq\{\bzero\}$, a contradiction to \eqref{secure_condition}. The lemma is proved.
\end{IEEEproof}

\begin{lemma}\label{lem_4}
Let $\vb_1^{\,(n-1)}$, $\vb_2^{\,(n-1)}$, $\cdots$, $\vb_{\w-1}^{\,(n-1)}$ be $\w-1$ linearly independent column $(n-1)$-vectors over a finite field $\Fq$, where $n=\w+r$. For each wiretap set $A\in \mA_r''$, define a set of column $(n-1)$-vectors as follows:
\begin{align}\label{mK_A}
\begin{split}
\mK_A=&\Big\{ \vl=\sum_{i=1}^{\w-1}\alpha_i\vb_i^{\,(n-1)}+\sum_{e\in A}\beta_e\vf_e^{\,(n-1)}:~\forall~\big(\alpha_i, 1\leq i \leq \w-1,\ \beta_e, e\in A\big)\in \Fq^{n-1}
\\
&\qquad\qquad\qquad\qquad\qquad\qquad\qquad\qquad\qquad \text{ s.t.  } \sum_{i=1}^{\w-1}\alpha_ib_{i,n}+\sum_{e\in A}\beta_ef_{e,n}=-1 \Big\},
\end{split}
\end{align}
where $\vf_e^{\,(n)}$, $e\in A$, are $r$ $n$-dimensional global encoding kernels of the $r$ edges in $A$, and $f_{e,n}$ is the last component of $\vf_e^{\,(n)}$. Then, for any column $(n-1)$-vector $\vk \in \Fq^{n-1}\setminus \bigcup_{A\in \mA_r''}\mK_A$, the following are satisfied:
\begin{itemize}
  \item the column $(n-1)$-vectors $\vb_i^{\,(n-1)}(\vk)$, $1\leq i \leq \w-1$, are linearly independent;
  \item $\mB_{\w-1}^{(n-1)}(\vk) \bigcap \mL_A^{(n-1)}(\vk)=\{ \bzero \}$, $\forall~A \in \mA_r''$.
  \end{itemize}
\end{lemma}
\begin{IEEEproof}
Let $A$ be an arbitrary edge subset in $\mA_r''$. In order to find a column vector $\vk\in \Fq^{n-1}$ such that $\vb_i^{\,(n-1)}(\vk)$, $1\leq i \leq \w-1$, are linearly independent and $\mB_{\w-1}^{(n-1)}(\vk) \bigcap \mL_A^{(n-1)}(\vk)=\{\bzero\}$, it suffices to find a column vector $\vk$ such that the $\w-1+r$ vectors $\vb_i^{\,(n-1)}(\vk)$, $1 \leq i \leq \w-1$, $\vf_e^{\,(n-1)}(\vk)$, $e \in A$, are linearly independent, or equivalently
\begin{align}\label{ineq-added}
\sum_{i=1}^{\w-1}\alpha_i\vb_i^{\,(n-1)}(\vk)+\sum_{e\in A}\beta_e\vf_e^{\,(n-1)}(\vk)\neq \bzero,
\end{align}
for any nonzero vector $\big( \alpha_i, 1\leq i \leq \w-1,\ \beta_e, e\in A \big) \in \Fq^{n-1}$ (here $n=\w+r$). Further, we write \eqref{ineq-added} as:
\begin{align*}
\sum_{i=1}^{\w-1}\alpha_i\Big( \vb_i^{\,(n-1)}+ b_{i,n}\vk \Big) +\sum_{e\in A}\beta_e \Big( \vf_e^{\,(n-1)}+f_{e,n}\vk \Big)\neq \bzero,
\end{align*}
or equivalently,
\begin{align}\label{lem_3-1}
-\Big(\sum_{i=1}^{\w-1}\alpha_ib_{i,n}+\sum_{e\in A}\beta_e f_{e,n}\Big)\cdot\vk \neq \sum_{i=1}^{\w-1}\alpha_i\vb_i^{\,(n-1)}+\sum_{e\in A}\beta_e\vf_e^{\,(n-1)}.
\end{align}

Based on \eqref{lem_3-1}, we consider the following two cases:

\noindent\underline{\bf\em Case 1:} Consider those nonzero vectors $\big(\alpha_i, 1\leq i \leq \w-1,\ \beta_e, e\in A \big)\in \Fq^{n-1}$ such that
\begin{align}\label{equ1_lem_4}
\sum_{i=1}^{\w-1}\alpha_ib_{i,n}+\sum_{e\in A}\beta_ef_{e,n}=0.
\end{align}
By~\eqref{equ1_lem_4}, the LHS in \eqref{lem_3-1} is always a zero vector. On the other hand, since $\big(\alpha_i, 1\leq i \leq \w-1,\ \beta_e, e\in A \big)$ is nonzero, and the vectors $\vb_i^{\,(n-1)}$, $1\leq i \leq \w-1$, $\vf_e^{\,(n-1)}$, $e \in A$, are linearly independent, the RHS in \eqref{lem_3-1} is nonzero. Thus, \eqref{lem_3-1} always holds for any $\vk$ in Case~1.

\noindent\underline{\bf\em Case 2:} Consider those nonzero vectors $\big(\alpha_i, 1\leq i\leq \w-1,\ \beta_e, e\in A \big)\in \Fq^{n-1}$ such that
\begin{align*}
\sum_{i=1}^{\w-1}\alpha_ib_{i,n}+\sum_{e\in A}\beta_e f_{e,n}\neq 0.
\end{align*}
Clearly, \eqref{lem_3-1} holds for Case~2 if and only if:
\begin{align*}%\label{2}
\vk\neq -\Big(\sum_{i=1}^{\w-1} \alpha_i b_{i,n}+\sum_{e\in A} \beta_e f_{e,n}\Big)^{-1}\cdot
\Big(\sum_{i=1}^{\w-1}\alpha_i\vb_i^{\,(n-1)}+\sum_{e\in A}\beta_e\vf_e^{\,(n-1)}\Big).
\end{align*}

Now we let
\begin{align}\label{mK_A_bar}
\begin{split}
\overline{\mK}_A=&\Big\{ \vl=-\left(\sum_{i=1}^{\w-1}\alpha_i b_{i,n}+\sum_{e\in A}\beta_e f_{e,n}\right)^{-1}\cdot
\left(\sum_{i=1}^{\w-1}\alpha_i\vb_i^{\,(n-1)}+\sum_{e\in A}\beta_e\vf_e^{\,(n-1)}\right):
\\
&\quad\qquad \forall~\big(\alpha_i, 1\leq i \leq \w-1,\ \beta_e, e\in A\big)\in \Fq^{n-1}\setminus\{\bzero\} \text{ s.t. } \sum_{i=1}^{\w-1}\alpha_ib_{i,n}+\sum_{e\in A}\beta_ef_{e,n}\neq0 \Big\}.
\end{split}
\end{align}
Combining the two cases above, we have proved that for any $\vk\in \Fq^{n-1}\setminus \overline{\mK}_A$, the $\w-1$ column $(n-1)$-vectors $\vb_i^{\,(n-1)}(\vk)$, $1\leq i \leq \w-1$, are linearly independent, and
$\mB_{\w-1}^{(n-1)}(\vk)\bigcap \mL_A^{(n-1)}(\vk)=\{\bzero\}$. Upon proving that $\overline{\mK}_A=\mK_A$ and considering all the edge subsets $A\in \mA_r''$, the lemma is proved.

We now prove that $\overline{\mK}_A=\mK_A$ (cf.~\eqref{mK_A} for $\mK_A$). Clearly, we have $\mK_A\subseteq \overline{\mK}_A$. To prove $\overline{\mK}_A\subseteq \mK_A$, we consider any column $(n-1)$-vector $\vl$ in $\overline{\mK}_A$. For this $\vl$, there exists a row $(n-1)$-vector $\big(\alpha_i, 1\leq i \leq \w-1,\ \beta_e, e\in A \big)$ in $\Fq^{n-1}$ such that $\sum_{i=1}^{\w-1}\alpha_ib_{i,n}+\sum_{e\in A}\beta_e f_{e,n}\neq 0$ and \begin{align*}
\vl=-\left(\sum_{i=1}^{\w-1}\alpha_i b_{i,n}+\sum_{e\in A}\beta_e f_{e,n}\right)^{-1}\cdot
\left(\sum_{i=1}^{\w-1}\alpha_i\vb_i^{\,(n-1)}+\sum_{e\in A}\beta_e\vf_e^{\,(n-1)}\right).
\end{align*}
Let
\begin{align*}
\alpha_i'=-\left(\sum_{i=1}^{\w-1}\alpha_i b_{i,n}+\sum_{e\in A}\beta_e f_{e,n}\right)^{-1}\cdot\alpha_i, \quad 1\leq i \leq \w-1,
\end{align*}
and
\begin{align*}
\beta_e'=-\left(\sum_{i=1}^{\w-1}\alpha_i b_{i,n}+\sum_{e\in A}\beta_e f_{e,n}\right)^{-1}\cdot\beta_e, \quad \forall~e\in A.
\end{align*}
Then we have
\begin{align*}
\vl=\sum_{i=1}^{\w-1}\alpha'_i\vb_i^{\,(n-1)}+\sum_{e\in A}\beta'_e\vf_e^{\,(n-1)},
\end{align*}
and
\begin{align*}
\sum_{i=1}^{\w-1}\alpha_i' b_{i,n}+\sum_{e\in A}\beta_e' f_{e,n}=-\left(\sum_{i=1}^{\w-1}\alpha_i b_{i,n}+\sum_{e\in A}\beta_e f_{e,n}\right)^{-1}\cdot\left(\sum_{i=1}^{\w-1}\alpha_i b_{i,n}+\sum_{e\in A}\beta_e f_{e,n}\right)=-1.
\end{align*}
This implies $\vl\in \mK_A$, proving that $\overline{\mK}_A\subseteq \mK_A$. We thus have proved that $\overline{\mK}_A=\mK_A$.
\end{IEEEproof}

The following theorem gives the prescription for designing the vector $\vk$.

\begin{thm}\label{thm_indep}
Let $\vb_1^{\,(n-1)}$, $\vb_2^{\,(n-1)}$, $\cdots$, $\vb_{\w-1}^{\,(n-1)}$ be $\w-1$ linearly independent column $(n-1)$-vectors over a finite field $\Fq$ of order $q>|\mA_r|$, where $n=\w+r$. Then, the set
\begin{align}\label{thm_indep-vk_choosen}
\Fq^{n-1}\setminus \Bigg[ \bigcup_{A\in \mA_r'} \Big(\mB_{\w-1}^{(n-1)}+\mL_A^{(n-1)}\Big)\bigcup_{A\in \mA_r''}\mK_A \Bigg]
\end{align}
is nonempty, and for any column $(n-1)$-vector $\vk$ in this set, the following are satisfied:
\begin{itemize}
  \item the column $(n-1)$-vectors $\vb_i^{\,(n-1)}(\vk)$, $1\leq i \leq \w-1$, are linearly independent;
  \item $\mB_{\w-1}^{(n-1)}(\vk)\bigcap \mL_A^{(n-1)}(\vk)=\{\bzero\}$, $\forall~A\in \mA_r$, i.e., the security condition in \eqref{secure_condition_2}.
\end{itemize}
\end{thm}

\begin{IEEEproof}
We only need prove that the set \eqref{thm_indep-vk_choosen} is nonempty if the field size $q>\lvert \mA_r \rvert$. The rest of the theorem are immediate consequences of Lemmas~\ref{lem_3} and \ref{lem_4}.

First, for an edge subset $A\in \mA_r'$, the $n-1$ column $(n-1)$-vectors $\vb_i^{\,(n-1)}$, $1\leq i \leq \w-1$, $\vf_e^{\,(n-1)}$, $e\in A$, are linearly dependent. Then we have
\begin{align*}
\dim\big(\mB_{\w-1}^{(n-1)}+\mL_A^{(n-1)}\big)\leq (\w-1+r)-1=n-2,
\end{align*}
and consequently,
$$\Big|\mB_{\w-1}^{(n-1)}+\mL_A^{(n-1)}\Big|\leq q^{n-2}.$$

Now, for an edge subset $A$ in $\mA_r''$, we consider any two vectors
$\big(\alpha_i, 1\leq i \leq \w-1,\ \beta_e, e\in A \big)$ and
$\big(\alpha'_i, 1\leq i \leq \w-1,\ \beta'_e, e\in A \big)$ in $\Fq^{n-1}$ such that
\begin{align}\label{eq_coe_l_l'}
\sum_{i=1}^{\w-1}\alpha_ib_{i,n}+\sum_{e\in A}\beta_ef_{e,n}=-1 \quad  \text{ and }\quad \sum_{i=1}^{\w-1}\alpha'_ib_{i,n}+\sum_{e\in A}\beta'_ef_{e,n}=-1.
\end{align}
Let
\begin{align}\label{eq_vl}
\vl=\sum_{i=1}^{\w-1}\alpha_i\vb_i^{\,(n-1)}+\sum_{e\in A}\beta_e\vf_e^{\,(n-1)}
\end{align}
and
\begin{align}\label{eq_vl'}
\vl{\,'}=\sum_{i=1}^{\w-1}\alpha'_i\vb_i^{\,(n-1)}+\sum_{e\in A}\beta'_e\vf_e^{\,(n-1)}.
\end{align}
We will prove that $\vl=\vl{\,'}$ if and only if
\begin{align*}
\big(\alpha_i, 1\leq i \leq \w-1,\ \beta_e, e\in A \big)=\big(\alpha'_i, 1\leq i \leq \w-1,\ \beta'_e, e\in A \big).
\end{align*}

The ``if'' part is evident. For the ``only if'' part, since $\vl=\vl{\,'}$, it follows from \eqref{eq_vl} and \eqref{eq_vl'} that
\begin{align}\label{equ4}
\sum_{i=1}^{\w-1}\alpha_i\vb_i^{\,(n-1)}+\sum_{e\in A}\beta_e\vf_e^{\,(n-1)}=
\sum_{i=1}^{\w-1}\alpha'_i\vb_i^{\,(n-1)}+\sum_{e\in A}\beta'_e\vf_e^{\,(n-1)}.
\end{align}
Note that the $n-1$ column $(n-1)$-vectors $\vb_i^{\,(n-1)}$, $1\leq i \leq \w-1$, $\vf_e^{\,(n-1)}$, $e\in A$, are linearly independent (since $A\in \mA_r''$), and so they form a basis of the vector space $\Fq^{n-1}$. Therefore, by \eqref{equ4}, we
immediately obtain that $\big(\alpha_i, 1\leq i \leq \w-1,\ \beta_e, e\in A \big)$ and $\big(\alpha'_i, 1\leq i \leq \w-1,\ \beta'_e, e\in A \big)$ are equal.

Therefore, the cardinality of $\mK_A$ is equal to the number of the vectors $\big(\alpha_i, 1\leq i \leq \w-1,\ \beta_e, e\in A \big) \in \Fq^{n-1}$ such that
\begin{align}\label{equ1_thm_indep}
\sum_{i=1}^{\w-1}\alpha_ib_{i,n}+\sum_{e\in A}\beta_e f_{e,n}=-1,
\end{align}
that is, $|\mK_A|=q^{n-2}$.

Thus, we have proved that
\begin{align*}%\label{thm_indep-cases}
\begin{cases}
\big|\mB_{\w-1}^{(n-1)}+\mL_A^{(n-1)}\big|\leq q^{n-2}, & \forall~A\in \mA_r';
\\
|\mK_A|=q^{n-2}, & \forall~A\in \mA_r''.
\end{cases}
\end{align*}
So, if $q>|\mA_r|$, we have
\begin{align}
& \left|\Fq^{n-1}\setminus \left[ \bigcup_{A\in \mA_r'} \Big(\mB_{\w-1}^{(n-1)}+\mL_A^{(n-1)}\Big)\bigcup_{A\in \mA_r''}\mK_A \right] \right|\nonumber\\
& = \Big|\Fq^{n-1}\Big|-\left|\bigcup_{A\in \mA_r'} \Big(\mB_{\w-1}^{(n-1)}+\mL_A^{(n-1)}\Big)\bigcup_{A\in \mA_r''}\mK_A\right|\nonumber\\
& \geq  q^{n-1}-\Big(\sum_{A\in \mA_r'} q^{n-2}+\sum_{A\in \mA_r''} q^{n-2} \Big)\nonumber\\
& =q^{n-1}-|\mA_r|\cdot q^{n-2}\nonumber\\
& =q^{n-2}(q-|\mA_r|)\nonumber\\
& >  0.\nonumber
\end{align}
The theorem is proved.
\end{IEEEproof}

Now, by Theorem~\ref{thm_indep}, the column $(n-1)$-vector $\vk$ can be taken to be any vector in the set in \eqref{thm_indep-vk_choosen}. Let
\begin{align*}
Q^{(n-1)}(\vk)=\Big[ I_{n-1} \ \ \vk\, \Big]\cdot Q^{(n)} = \Big[ \vb_1^{\,(n-1)}(\vk) \ \ \vb_2^{\,(n-1)}(\vk) \ \ \cdots \ \ \vb_n^{\,(n-1)}(\vk) \Big].
\end{align*}
Then we have $\Rank\big(Q^{(n-1)}(\vk)\big)=n-1$ since $\Rank\big(Q^{(n)}\big)=n$. Furthermore, together with the linear independence of $\vb_1^{\,(n-1)}(\vk)$, $\vb_2^{\,(n-1)}(\vk)$, $\cdots$, $\vb_{\w-1}^{\,(n-1)}(\vk)$ from  Theorem~\ref{thm_indep}, there exist $r$ column vectors in the last $r+1$ column vectors of $Q^{(n-1)}(\vk)$, say $\vb_{\w}^{\,(n-1)}(\vk)$, $\vb_{\w+1}^{\,(n-1)}(\vk)$, $\cdots$, $\vb_{n-1}^{\,(n-1)}(\vk)$,  such that $\vb_1^{\,(n-1)}(\vk)$, $\vb_2^{\,(n-1)}(\vk)$, $\cdots$, $\vb_{n-1}^{\,(n-1)}(\vk)$ are linearly independent. Immediately, we obtain an $(n-1)\times(n-1)$ invertible matrix
\begin{align*}
Q^{(n-1)}=\begin{bmatrix} \vb_1^{\,(n-1)}(\vk) & \vb_2^{\,(n-1)}(\vk) & \cdots\ \vb_{n-1}^{\,(n-1)}(\vk) \end{bmatrix}.
\end{align*}

Therefore, we have constructed an $(n-1)$-dimensional SLNC
$$(Q^{(n-1)})^{-1}\cdot\, \mC_{n-1}= (Q^{(n-1)})^{-1}\cdot \left[ I_{n-1} \ \ \vk\, \right] \cdot \mC_n = \Big\{ (Q^{(n-1)})^{-1}\cdot\vf_e^{\,(n-1)}(\vk):~e\in E \Big\},$$
which not only achieves the information rate $\w-1$ and security level of $r$, but also has the same local encoding kernels as the original $n$-dimensional SLNC $(Q^{(n)})^{-1}\cdot\, \mC_{n}$  at all the non-source nodes.
%=\big\{ (Q^{(n)})^{-1} \cdot \vf^{\,(n)}_e:~e\in E \big\}$

\section{An Algorithm for Code Construction}\label{section_algorithm}

In the last section, we have presented an approach for designing a family of local-encoding-preserving SLNCs for a fixed security level and multiple rates. In particular, Theorem~\ref{thm_indep} gives the prescription for designing an appropriate vector $\vk$ that is crucial for constructing a local-encoding-preserving SLNC with rate reduced by one. However, Theorem~\ref{thm_indep} does not provide a method to find $\vk$ readily. The following Algorithm~\ref{algo-1} provides an efficient method to find $\vk$ and gives a polynomial-time implementation for constructing an $(n-1)$-dimensional SLNC with rate $\w-1$ and security level $r$ (here $n=\w+r$) from an $n$-dimensional SLNC with rate $\w$ and security level $r$, and the two SLNCs have the same local encoding kernel at each intermediate node.

\begin{figure}[!t]
\centering
\begin{tikzpicture}[%Define standard arrow tip
>=stealth',
%Define style for different line styles
help lines/.style={dotted, thick},
axis/.style={<->},
important line/.style={thick, -latex},
connection/.style={thick, dashed}]
    % Axis
    \coordinate (y) at (0,6);
    \coordinate (x) at (6,0);
    \draw[<->] (y) node[above] {$r$} -- (0,0) node [below left]{$0$} --  (x) node[right]
    {$\w$};
    \path
    coordinate (start) at (0,5)
    coordinate (p00) at (0,0)
    coordinate (p01) at (0,1)
    coordinate (p02) at (0,2)
    coordinate (p03) at (0,3)
    coordinate (p04) at (0,4)
    coordinate (p10) at (1,0)
    coordinate (p11) at (1,1)
    coordinate (p12) at (1,2)
    coordinate (p13) at (1,3)
    coordinate (p14) at (1,4)
    coordinate (p20) at (2,0)
    coordinate (p21) at (2,1)
    coordinate (p22) at (2,2)
    coordinate (p23) at (2,3)
    coordinate (p30) at (3,0)
    coordinate (p31) at (3,1)
    coordinate (p32) at (3,2)
    coordinate (p40) at (4,0)
    coordinate (p41) at (4,1)
    coordinate (end) at (5,0);
    \draw[help lines] (start) node [left] {$C_{\min}$} -- (end) node[below]
    {$C_{\min}$};
    \draw[help lines] (p14) node [left] {} -- (p04) node[below] {};
    \draw[help lines] (p23) node [left] {} -- (p03) node[below] {};

    \draw[important line] (p32) node [left] {} -- (p22) node[left] {};
    \draw[important line] (p22) node [left] {} -- (p12) node[left] {};
    \draw[important line] (p12) node [left] {} -- (p02) node[left] {$r$};

    \draw[help lines] (p41) node [left] {} -- (p01) node[below] {};
%    \draw[help lines] (end) node [left] {} -- (p00) node[below] {};
    \draw[help lines] (p10) -- (p14);
    \draw[help lines] (p20) -- (p23);
    \draw[help lines] (p30) -- (p32);
    \draw[help lines] (p40) -- (p41);
    %\draw[help lines] (p01) -- (p41);
    %\draw[help lines] (p02) -- (p32);
    %\draw[help lines] (p03) -- (p23);
    %\draw[help lines] (p04) -- (p14);
    % Help coordinates for drawing the curve
     \filldraw [black]
     (start) circle (1pt)
     (p00) circle (1pt)
     (p01) circle (1pt)
     (p02) circle (1pt)
     (p03) circle (1pt)
     (p04) circle (1pt)
     (p10) circle (1pt)
     (p11) circle (1pt)
     (p12) circle (1pt)
     (p13) circle (1pt)
     (p14) circle (1pt)
     (p20) circle (1pt)
     (p21) circle (1pt)
     (p22) circle (1pt)
     (p23) circle (1pt)
     (p30) circle (1pt)
     (p31) circle (1pt)
     (p32) circle (1pt)
     (p40) circle (1pt)
     (p41) circle (1pt)
     (end) circle (1pt);
     %(slut) circle (1pt)
    %\filldraw [black]
\end{tikzpicture}
\caption{Local-encoding-preserving SLNCs for a fixed security-level and a flexible rate.}
\label{fig-fix-SecLevel-flexible-Rate}
\end{figure}
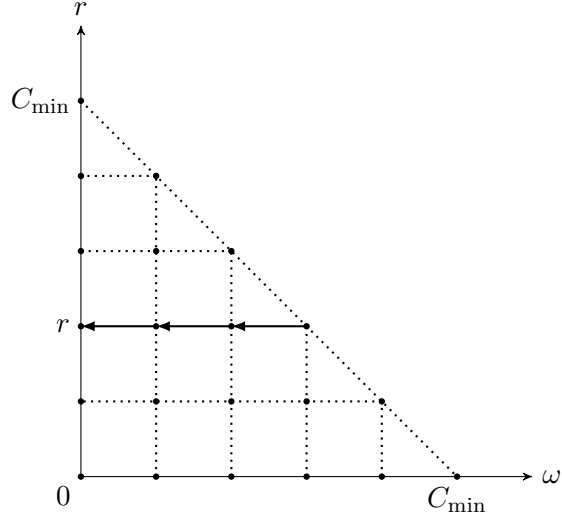

Starting with an SLNC with a security level $r$ and accordingly the maximum rate $C_{\min}-r$ and then using Algorithm~\ref{algo-1} repeatedly, we can obtain a family of local-encoding-preserving SLNCs with the fixed security level $r$ and rates from $C_{\min}-r$ to $0$. This procedure is illustrated in Fig.~\ref{fig-fix-SecLevel-flexible-Rate}.

\begin{algorithm}[!htp]\small
%\dontprintsemicolon
\SetAlgoLined
%\SetLine
\KwIn{An $n$-dimensional linear network code $\mC_n$ of global encoding kernels $\vf_e^{\,(n)},~e\in E$ over a finite field $\Fq$ of order $q>|\mA_r|$, and an $\Fq$-valued $n \times n$ invertible matrix $Q^{(n)}=\Big[\vb_1^{\,(n)} \ \vb_2^{\,(n)} \ \cdots \ \vb_n^{\,(n)} \Big]$ such that $(Q^{(n)})^{-1}\cdot\,\mC_n$ is an $n$-dimensional SLNC with rate $\w$ and security level $r$, where $n=\w+r$.}
\KwOut{Matrices $Q^{(n-1)}$ and $K_s^{(n-1)}(\vk)$ corresponding to the linear encoding operation and the local encoding kernel at the source node, respectively.\newline
\tcp*[f]{\rm\footnotesize The linear encoding operation corresponding to $Q^{(n-1)}$ at the source node, the updated local encoding kernel $K_s^{(n-1)}(\vk)$ at the source node, and the unchanged local encoding kernels at all the non-source nodes together constitute an $(n-1)$-dimensional SLNC of rate $\w-1$ and security level $r$.}}
\BlankLine
\Begin{
\nl choose $\w-1$ linearly independent vectors in $\vb_1^{\,(n-1)}$, $\vb_2^{\,(n-1)}$,$\cdots$, $\vb_{\w}^{\,(n-1)}$, say $\vb_1^{\,(n-1)}$, $\vb_2^{\,(n-1)}$,$\cdots$, $\vb_{\w-1}^{\,(n-1)}$ without loss of generality\;
\nl partition $\mA_r$ into $\mA_r'$ and $\mA_r''$\;
\nl choose a column $(n-1)$-vector $\vh \in \Fq^{n-1}\setminus \bigcup_{A\in \mA_r'} \big(\mB_{\w-1}^{(n-1)}+\mL_A^{(n-1)}\big)$\;
\nl      \For{ each $A\in \mA_r''$ }{
\nl      calculate the unique row $(n-1)$-vector $\big(\alpha_i, 1\leq i \leq \w-1,\ \beta_e, e\in A \big)\in \Fq^{n-1}$ such that\newline
      \qquad $\vh=\sum_{i=1}^{\w-1}\alpha_i\vb_i^{\,(n-1)}+\sum_{e\in A} \beta_e\vf_e^{\,(n-1)}$\;
\nl   compute $\theta_A=\sum_{i=1}^{\w-1}\alpha_i b_{i,n}+\sum_{e\in A}\beta_e f_{e,n}$\;}
\nl choose a nonzero element $\theta$ in $\Fq$ such that $\theta\cdot\theta_A\neq -1,~\forall~A\in \mA_r''$\;
\nl calculate the vector $\vk=\theta\vh$;
\tcp*[f]{\rm\footnotesize $\vk\in \Fq^{n-1}\setminus \big[\bigcup_{A\in \mA_r'} \big(\mB_{\w-1}^{(n-1)}+\mL_A^{(n-1)}\big)\bigcup_{A\in \mA_r''}\mK_A\big]$.}\newline
\nl compute
$K_s^{(n-1)}(\vk)=\Big[ I_{n-1} \ \ \vk\, \Big]\cdot K_s$;
\tcp*[f]{\rm\footnotesize $K_s^{(n-1)}(\vk)=\Big[ \vf_e^{\,(n-1)}(\vk):~e\in \Out(s) \Big]$.}\newline
\nl compute $Q^{(n-1)}(\vk)=\Big[ I_{n-1} \ \ \vk\, \Big]\cdot Q^{(n)}$; \tcp*[f]{\rm\footnotesize $Q^{(n-1)}(\vk)=\Big[\vb_1^{\,(n-1)}(\vk) \ \vb_2^{\,(n-1)}(\vk) \ \cdots \ \vb_{n}^{\,(n-1)}(\vk) \Big]$.}\newline
\nl remove a column vector from the last $r+1$ (here $r=n-\w$) vectors of $Q^{(n-1)}(\vk)$, i.e., $\vb_{i}^{\,(n-1)}(\vk)$, $\w \leq i \leq n$, such that the remaining $(n-1)\times (n-1)$ matrix, denoted by $Q^{(n-1)}$, is invertible\;
\nl return $Q^{(n-1)}$ and $K_s^{(n-1)}(\vk)$.\newline
\tcp*[f]{\rm\footnotesize Now, $(Q^{(n-1)})^{-1}\cdot\,\mC_{n-1}$ is an $\Fq$-valued SLNC with security level $r$ and rate $\w-1$, and has the same local encoding kernels as $(Q^{(n)})^{-1}\cdot\,\mC_n$ at all the non-source nodes. However, note that the calculation of all the global encoding kernels  of $\mC_{n-1}$, i.e., $\vf_e^{\,(n-1)}(\vk)$, $e\in E$, is not necessary. Instead, we only need to compute $K_s^{(n-1)}(\vk)$, the new local encoding kernel at the source node $s$, and continue to use the original local encoding coefficients at each non-source node for encoding.}
}
\caption{Construction of a rate-$(\w-1)$ and security-level-$r$ SLNC from a rate-$\w$ and security-level-$r$ SLNC, both of which have the same local encoding kernels at all the non-source nodes.}
\label{algo-1}
\end{algorithm}

\newpage

\noindent\textbf{Verification of Algorithm~\ref{algo-1}:}

For the purpose of verifying Algorithm~\ref{algo-1}, it suffices to verify that the column $(n-1)$-vector $\vk$ chosen in Line~8 is in the set in \eqref{thm_indep-vk_choosen}, i.e.,
\begin{align}\label{eq_verific_algo-1}
\vk\in \Fq^{n-1}\setminus \Bigg[ \bigcup_{A\in \mA_r'} \Big(\mB_{\w-1}^{(n-1)}+\mL_A^{(n-1)}\Big)\bigcup_{A\in \mA_r''}\mK_A \Bigg].
\end{align}

First, let $\vh$ be the column $(n-1)$-vector in $\Fq^{n-1}\setminus \bigcup_{A\in \mA_r'} \big(\mB_{\w-1}^{(n-1)}+\mL_A^{(n-1)}\big)$ that has been chosen in Line~5. Note that
\begin{align}\label{vk-1st-chosen}
\vk=\theta\vh \in \Fq^{n-1}\setminus \bigcup_{A\in \mA_r'} \big(\mB_{\w-1}^{(n-1)}+\mL_A^{(n-1)}\big),
\end{align}
where $\theta$ is as chosen in Line~7.

Now, consider any wiretap set $A\in\mA_r''$. It follows that $\vb_i^{\,(n-1)}$, $1\leq i \leq \w-1$, $\vf_e^{\,(n-1)}$, $e\in A$, are $n-1$ linearly independent column $(n-1)$-vectors, and thus form a basis of the vector space $\Fq^{n-1}$. So there exists the unique row $(n-1)$-vector $\big(\alpha_{i}, 1\leq i \leq \w-1,\ \beta_{e}, e\in A \big)$ such that
\begin{align*}
\vh=\sum_{i=1}^{\w-1}\alpha_i\vb_i^{\,(n-1)}+\sum_{e\in A} \beta_e\vf_e^{\,(n-1)}.
\end{align*}

%Next, we enhance a result we have proved in the proof of Theorem~\ref{thm_indep}, namely that, for any two nonzero vectors
%$\big(\alpha_{1,i}, 1\leq i \leq \w-1,\ \beta_{1,e}, e\in A \big)$ and
%$\big(\alpha_{2,i}, 1\leq i \leq \w-1,\ \beta_{2,e}, e\in A \big)$ in $\Fq^{n-1}$ such that
%\begin{align*}
%\sum_{i=1}^{\w-1}\alpha_{1,i}b_{i,n}+\sum_{e\in A}\beta_{1,e}f_{e,n}=-1 \quad  \text{ and }\quad \sum_{i=1}^{\w-1}\alpha_{2,i}b_{i,n}+\sum_{e\in A}\beta_{2,e}f_{e,n}=-1,
%\end{align*}
%$\vl_{1}=\vl_2$ if and only if $\big(\alpha_{1,i}, 1\leq i \leq \w-1,\ \beta_{1,e}, e\in A \big)=\big(\alpha_{2,i}, 1\leq i \leq \w-1,\ \beta_{2,e}, e\in A \big)$, where
%\begin{align*}
%\vl_{1}=\sum_{i=1}^{\w-1}\alpha_{1,i}\vb_i^{\,(n-1)}+\sum_{e\in A}\beta_{1,e}\vf_e^{\,(n-1)}
%\end{align*}
%and
%\begin{align*}
%\vl_{2}=\sum_{i=1}^{\w-1}\alpha_{2,i}\vb_i^{\,(n-1)}+\sum_{e\in A}\beta_{2,e}\vf_e^{\,(n-1)}.
%\end{align*}
%In fact, if $\big(\alpha_{1,i}, 1\leq i \leq \w-1,\ \beta_{1,e}, e\in A \big)\neq\big(\alpha_{2,i}, 1\leq i \leq \w-1,\ \beta_{2,e}, e\in A \big)$, we see that $\big(\alpha_{1,i}, 1\leq i \leq \w-1,\ \beta_{1,e}, e\in A \big)$ and $\big(\alpha_{2,i}, 1\leq i \leq \w-1,\ \beta_{2,e}, e\in A \big)$ are linearly independent. Together with the linear independence of $\vb_i^{\,(n-1)}$, $1\leq i \leq \w-1$, $\vf_e^{\,(n-1)}$, $e\in A$, this further implies that $\vl_{1}$ and $\vl_{2}$ are linearly independent.

We prove in the following that the vector $\vk=\theta\vh$ calculated in Line~8 is not equal to any vector $\vl{\;'}\in \bigcup_{A\in \mA_r''}\mK_A$. Fix any $A\in \mA_r''$ and consider any $\vl{\;'} \in \mK_A$. Let
\begin{align*}
\vl{\,'}=\sum_{i=1}^{\w-1}\alpha'_i\vb_i^{\,(n-1)}+\sum_{e\in A} \beta'_e\vf_e^{\,(n-1)},
\end{align*}
where $\big(\alpha_{i}', 1\leq i \leq \w-1,\ \beta_{e}', e\in A \big)$ is a row $(n-1)$-vector such that
\begin{align}\label{equ-verify-algo1}
\sum_{i=1}^{\w-1}\alpha_{i}'b_{i,n}+\sum_{e\in A}\beta_{e}' f_{e,n}=-1.
\end{align}
Note that $\big(\alpha_{i}', 1\leq i \leq \w-1,\ \beta_{e}', e\in A \big)$ is unique because $\vb_i^{\,(n-1)}$, $1\leq i \leq \w-1$, $\vf_e^{\,(n-1)}$, $e\in A$, form a basis of $\Fq^{n-1}$ (since $A\in \mA_r''$).

\noindent\underline{\bf\em Case 1:} $\big(\alpha_{i}', 1\leq i \leq \w-1,\ \beta_{e}', e\in A \big)$ and $\big(\alpha_{i}, 1\leq i \leq \w-1,\ \beta_{e}, e\in A \big)$ are linearly independent. By the linear independence of $\vb_i^{\,(n-1)}$, $1\leq i \leq \w-1$, $\vf_e^{\,(n-1)}$, $e\in A$, the column $(n-1)$-vectors $\vh$ and $\vl{\,'}$ are linearly independent, and immediately we have $\vk=\theta\vh\neq \vl{\,'}$ for any nonzero element $\theta$ in $\Fq$.

\noindent\underline{\bf\em Case 2:} $\big(\alpha'_{i}, 1\leq i \leq \w-1,\ \beta'_{e}, e\in A \big)$ and $\big(\alpha_{i}, 1\leq i \leq \w-1,\ \beta_{e}, e\in A \big)$ are linearly dependent. Let
\begin{align*}
\theta_A= \sum_{i=1}^{\w-1}\alpha_{i}b_{i,n}+\sum_{e\in A}\beta_{e} f_{e,n}.
\end{align*}
We now prove $\vk=\theta\vh\neq \vl{\;'}$, where $\theta$ is as chosen in Line~7. Note that
\begin{align*}
\sum_{i=1}^{\w-1}(\theta\alpha_{i})\;b_{i,n}+\sum_{e\in A}(\theta\beta_{e})\; f_{e,n}=\theta\cdot\left(\sum_{i=1}^{\w-1}\alpha_{i}b_{i,n}+\sum_{e\in A}\beta_{e} f_{e,n} \right)=\theta\cdot\theta_A\neq -1.
\end{align*}
Together with \eqref{equ-verify-algo1}, this implies that $\theta\big(\alpha_{i}, 1\leq i \leq \w-1,\ \beta_{e}, e\in A \big)$ and $\big(\alpha'_{i}, 1\leq i \leq \w-1,\ \beta'_{e}, e\in A \big)$ are not equal (but linearly dependent). By the linear independence of $\vb_i^{\,(n-1)}$, $1\leq i \leq \w-1$, $\vf_e^{\,(n-1)}$, $e\in A$, this further implies
\begin{align*}
\theta\vh=\sum_{i=1}^{\w-1}(\theta\alpha_i)\vb_i^{\,(n-1)}+\sum_{e\in A} (\theta\beta_e)\vf_e^{\,(n-1)}\neq \sum_{i=1}^{\w-1}\alpha'_i\vb_i^{\,(n-1)}+\sum_{e\in A} \beta'_e\vf_e^{\,(n-1)}=\vl{\;'}.
\end{align*}

Combining the above two cases, we have proved that $\vk\notin \mK_A$ for all $A\in\mA_r''$. Together with \eqref{vk-1st-chosen}, we have verified that $\vk$ is in the set in \eqref{thm_indep-vk_choosen}.

We now give an example to illustrate Algorithm~\ref{algo-1}.

\begin{eg}\label{eg-1}

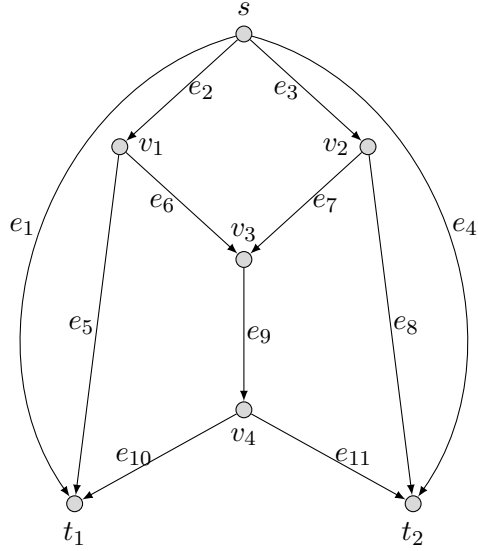
\begin{figure}[!h]
\tikzstyle{vertex}=[draw,circle,fill=gray!30,minimum size=6pt, inner sep=0pt]
  \centering
{
 \begin{tikzpicture}[x=0.6cm]
    \draw (0,0) node[vertex]     (s)[label=above:$s$] {};
    \draw (-2.75,-1.5) node[vertex] (1) [label=right:$v_1$] {};
    \draw ( 2.75,-1.5) node[vertex] (2) [label=left:$v_2$] {};
    \draw ( 0,-3) node[vertex] (3) [label=above:$v_3$] {};
    \draw ( 0,-5) node[vertex] (6) [label=below:$v_4$] {};
    \draw (-3.75,-6.25) node[vertex] (4) [label=below:$t_1$] {};
    \draw ( 3.75,-6.25) node[vertex] (5) [label=below:$t_2$] {};

    \draw[->,>=latex] (s) -- (1) node[midway, auto,swap, right=-0.5mm] {$e_2$};
    \draw[->,>=latex] (s) -- (2) node[midway, auto, left=-0.5mm] {$e_3$};
    \draw[->,>=latex] (1) -- (4) node[midway, auto,swap, left=-1mm] {$e_5$};
    \draw[->,>=latex] (1) -- (3) node[midway, auto,left=-0.5mm] {$e_6$};
    \draw[->,>=latex] (2) -- (3) node[midway, auto,right=-0.5mm] {$e_7$};
    \draw[->,>=latex] (2) -- (5) node[midway, auto, right=-1mm] {$e_8$};
    \draw[->,>=latex] (3) -- (6) node[midway, auto, right=-1mm] {$e_9$};
    \draw[->,>=latex] (6) -- (4) node[midway, auto, left=-0.5mm] {$e_{10}$};
    \draw[->,>=latex] (6) -- (5) node[midway, auto,swap, right=-0.5mm] {$e_{11}$};
    \draw[->,>=latex] (s) edge[bend right=55] node[pos=0.5,left=-0.5mm] {$e_1$} (4) ;
    \draw[->,>=latex] (s) edge[bend left=55]  node[pos=0.5,right=-0.5mm] {$e_4$} (5);
    \end{tikzpicture}
}
\caption{The network $G=(V,E)$.}
  \label{fig}
\end{figure}

Consider the network $G=(V,E)$ depicted in Fig.~\ref{fig}. Let
\begin{align}\label{local_kernels_EX}
\mC_3=\left\{ K_{s}=\left[\begin{smallmatrix} 0 & 1 & 1 & 0 \\ 1 & 0 & 0 & 1 \\ 1 & 1 & 2 & 2 \end{smallmatrix}\right],\quad
K_{1}=K_{2}=K_{4}=\left[\begin{smallmatrix} 1 & 1 \end{smallmatrix}\right],\quad K_{3}=\big[\begin{smallmatrix} 4 \\ 1 \end{smallmatrix}\big] \right\}
\end{align}
be a $3$-dimensional linear network code over the field $\mathbb{F}_5$ on $G$, where we use $K_{s}$ and $K_{i}$ to denote the local encoding kernels at the source node $s$ and the intermediate nodes $v_i$, $1\leq i \leq 4$, respectively. We use $\vf_i^{\,(3)}$ to denote the global encoding kernels of the edges $e_i$ for all $1\leq i \leq 11$, which according to~\eqref{equ_ext_f} are calculated as follows:
\begin{align}\label{global_kernels_EX}
\begin{split}
&\vf_1^{\,(3)}=\left[\begin{smallmatrix} 0 \\ 1 \\ 1 \end{smallmatrix}\right], \quad \vf_2^{\,(3)}=\vf_5^{\,(3)}=\vf_6^{\,(3)}=\left[\begin{smallmatrix} 1 \\ 0 \\ 1 \end{smallmatrix}\right], \quad \vf_3^{\,(3)}=\vf_7^{\,(3)}=\vf_8^{\,(3)}=\left[\begin{smallmatrix} 1 \\ 0 \\ 2 \end{smallmatrix}\right],\\
&\vf_4^{\,(3)}=\left[\begin{smallmatrix} 0 \\ 1 \\ 2 \end{smallmatrix}\right],\quad  \vf_9^{\,(3)}=\vf_{10}^{\,(3)}=\vf_{11}^{\,(3)}=\left[\begin{smallmatrix} 0 \\ 0 \\ 1 \end{smallmatrix}\right].
\end{split}
\end{align}

Consider rate $\w=2$ and security level $r=1$. The set of the primary edge subsets of size $1$ is
\begin{align*}
\mA_1=\big\{ \{e_1\}, \{e_2\}, \{e_3\}, \{e_4\}, \{e_9\}\big\}.
\end{align*}
Let
\begin{align*}
Q^{(3)}=\begin{bmatrix}\vb_1^{\,(3)} & \vb_2^{\,(3)} & \vb_3^{\,(3)} \end{bmatrix}=\left[\begin{smallmatrix}1 & 0 & 0 \\ 0 & 1 &0 \\ 0 & 0 & 1 \end{smallmatrix}\right],
\end{align*}
which is clearly invertible. Furthermore, we see that the following are satisfied:
\begin{align*}
\big\langle  \vb_1^{\,(3)},~\vb_2^{\,(3)} \big\rangle \bigcap
\big\langle \vf_i^{\,(3)} \big\rangle=\{\bzero\},\quad \forall~i=1,2,3,4,9,\quad \text{i.e.,}\quad \forall~A\in \mA_1.
\end{align*}
Thus, $(Q^{(3)})^{-1}\cdot\mC_3=\mC_3$ is an $\mathbb{F}_5$-valued $3$-dimensional SLNC with security level $1$ and rate $2$.

In the following, we will use Algorithm~\ref{algo-1} to construct an $\mathbb{F}_5$-valued SLNC with the fixed security level $1$ and a lower rate $\w-1=1$, which also has the same local encoding kernels as the SLNC $(Q^{(3)})^{-1}\cdot\,\mC_3$ at all the non-source nodes. Note that $\vb_1^{\,(2)}=\big[\begin{smallmatrix} 1 \\ 0 \end{smallmatrix}\big]$ is nonzero and so itself is linearly independent. By Algorithm~\ref{algo-1}, we partition $\mA_1$ into the following two disjoint subsets $\mA_1'$ and $\mA_1''$:
\begin{align*}
\mA_1'&=\big\{ \{e_i\} \in \mA_1:~\vb_1^{\,(2)},~\vf_i^{\,(2)}~\text{ are linearly dependent} \big\}=\big\{\{e_2\}, \{e_3\}, \{e_9\} \big\},\\
\mA_1''&=\big\{ \{e_i\} \in \mA_1:~\vb_1^{\,(2)},~\vf_i^{\,(2)}~\text{ are linearly independent} \big\}=\big\{ \{e_1\}, \{e_4\} \big\}.
\end{align*}

First, for each $A\in \mA_1'$, note that $\mL_A^{(n-1)}=\mL_A^{(2)}$ (cf.~\eqref{L_A}) is either a $1$-dimensional subspace spanned by $\big[\begin{smallmatrix} 1 \\ 0 \end{smallmatrix}\big]$ ($\mL_{\{ e_2 \}}^{(2)}$ and $\mL_{\{ e_3 \}}^{(2)}$) or a null space ($\mL_{\{ e_9 \}}^{(2)}$). Thus, we arbitrarily choose an $\mathbb{F}_5$-valued column $2$-vector
\begin{align*}
\vh \in
\Fq^{n-1}\setminus \bigcup_{A\in \mA_r'} \big(\mB_{\w-1}^{(n-1)}+\mL_A^{(n-1)}\big) =\mathbb{F}_5^2\setminus \big\langle \big[\begin{smallmatrix} 1 \\ 0 \end{smallmatrix}\big] \big\rangle, \quad \text{ say}\quad \vh=\begin{bmatrix} 0 \\ 1 \end{bmatrix}.
\end{align*}

Next, for all $A\in \mA_1''$, i.e., $\{e_1\}$ and $\{e_4\}$, we respectively calculate
\begin{align*}
\vh=\begin{bmatrix} 0 \\ 1 \end{bmatrix}=\alpha_{e_1}\vb_1^{\,(2)}+\beta_{e_1}\vf_1^{\,(2)}
=0\cdot\begin{bmatrix} 1\\ 0 \end{bmatrix}+1\cdot\begin{bmatrix} 0 \\ 1 \end{bmatrix},
\quad \text{ i.e., }\quad (\alpha_{e_1}, \beta_{e_1})=(0,1),
\end{align*}
and
\begin{align*}
\vh=\begin{bmatrix} 0 \\ 1 \end{bmatrix}=\alpha_{e_4}\vb_1^{\,(2)}+\beta_{e_4}\vf_4^{\,(2)}
=0\cdot\begin{bmatrix} 1\\ 0 \end{bmatrix}+1\cdot\begin{bmatrix} 0 \\ 1 \end{bmatrix},
\quad \text{ i.e., }\quad (\alpha_{e_4}, \beta_{e_4})=(0,1).
\end{align*}
According to $(\alpha_{e_1}, \beta_{e_1})=(0,1)$ and $(\alpha_{e_4}, \beta_{e_4})=(0,1)$, we calculate
\begin{align*}
\theta_{\{e_1\}}=\alpha_{e_1}b_{1,3}+\beta_{e_1}f_{e_1,3}=1\cdot1=1,\quad
\text{ and }\quad
\theta_{\{e_4\}}=\alpha_{e_4}b_{1,3}+\beta_{e_4}f_{e_4,3}=1\cdot2=2.
\end{align*}
Consequently, we choose $\theta=3$ (since $\theta\cdot\theta_{\{e_1\}}=3\neq -1$ and $\theta\cdot\theta_{\{e_4\}}=1\neq -1$), and let \begin{align*}
\vk=\theta\cdot\vh=\left[\begin{matrix} 0 \\ 3 \end{matrix}\right].
\end{align*}

In fact, the sets $\mK_{\{e_1\}}$ and $\mK_{\{e_4\}}$ (see~\eqref{mK_A}) are
\begin{align*}
\mK_{\{e_1\}}&=\Big\{
\big[\begin{smallmatrix} 0 \\ 4 \end{smallmatrix}\big],
\big[\begin{smallmatrix} 4 \\ 4 \end{smallmatrix}\big],
\big[\begin{smallmatrix} 3 \\ 4 \end{smallmatrix}\big],
\big[\begin{smallmatrix} 2 \\ 4 \end{smallmatrix}\big],
\big[\begin{smallmatrix} 1 \\ 4 \end{smallmatrix}\big] \Big\},\\
\mK_{\{e_4\}}&=\Big\{
\big[\begin{smallmatrix} 0 \\ 2 \end{smallmatrix}\big],
\big[\begin{smallmatrix} 2 \\ 2 \end{smallmatrix}\big],
\big[\begin{smallmatrix} 4 \\ 2 \end{smallmatrix}\big],
\big[\begin{smallmatrix} 1 \\ 2 \end{smallmatrix}\big],
\big[\begin{smallmatrix} 3 \\ 2 \end{smallmatrix}\big]\Big\},
\end{align*}
and thus we see that
\begin{align*}
\vk=\begin{bmatrix} 0 \\ 3 \end{bmatrix} \in
\Fq^{n-1}\setminus \left[ \bigcup_{A\in \mA_r'} \big(\mB_{\w-1}^{(n-1)}+\mL_A^{(n-1)}\big)\bigcup_{A\in \mA_r''}\mK_A \right]
=\mathbb{F}_5^2\setminus \Big[ \big\langle \big[\begin{smallmatrix} 1 \\ 0 \end{smallmatrix}\big] \big\rangle \bigcup \mK_{\{e_1\}} \bigcup \mK_{\{e_4\}} \Big].
\end{align*}

With the vector $\vk$ we have chosen, we compute
\begin{align*}
K_s^{(2)}(\vk)&=\Big[ I_{2} \ \ \vk\, \Big]\cdot K_s=\Big[ \vf_1^{\,(2)}(\vk)\ \ \vf_2^{\,(2)}(\vk)\ \ \vf_3^{\,(2)}(\vk)\ \ \vf_4^{\,(2)}(\vk) \Big]=\big[\begin{smallmatrix} 0 & 1 & 1 & 0 \\ 4 & 3 & 1 & 2 \end{smallmatrix}\big],\\
Q^{(2)}(\vk)&=\Big[ I_{2} \ \ \vk\, \Big]\cdot Q^{(3)}
=\Big[\vb_1^{\,(2)}(\vk)\ \ \vb_2^{\,(2)}(\vk)\ \ \vb_{3}^{\,(2)}(\vk) \Big]
=\big[\begin{smallmatrix} 1 & 0 & 0 \\ 0 & 1 & 3 \end{smallmatrix}\big].
\end{align*}
We further let $Q^{(2)}=\Big[\vb_1^{\,(2)}(\vk) \ \ \vb_2^{\,(2)}(\vk)\Big]=I_2$. Then, we obtain a $2$-dimensional SLNC $(Q^{(2)})^{-1}\cdot\,\mC_2$ with security level $1$ and rate $1$, where
\begin{align*}
\mC_2=\Big[ I_{2} \ \ \vk\, \Big]\cdot\mC_3=
\left\{ K_s^{(2)}(\vk)=\big[\begin{smallmatrix} 0 & 1 & 1 & 0 \\ 4 & 3 & 1 & 2 \end{smallmatrix}\big],\quad
K_{1}=K_{2}=K_{4}=\left[\begin{smallmatrix} 1 & 1 \end{smallmatrix}\right],\quad K_{3}=\big[\begin{smallmatrix} 4 \\ 1 \end{smallmatrix}\big] \right\}
\end{align*}
(cf.~Theorem~\ref{thm_LNC_transformation}). Clearly, $(Q^{(2)})^{-1}\cdot\,\mC_2$ has the same local encoding kernels as $(Q^{(3)})^{-1}\cdot\,\mC_3$ at all the non-source nodes. In the use of $(Q^{(2)})^{-1}\cdot\,\mC_2$, {\rm\rmnum{1})} at the source node $s$, we use $Q^{\,(2)}$ to linearly encode the source message and the key and then use $K_s^{(2)}(\vk)$ to encode the linearly-encoded message at the source node $s$; and then {\rm\rmnum{2})} at each the intermediate node $v_1,v_2,v_3,v_4$, use the unchanged local encoding kernel for encoding.
\end{eg}

\noindent\textbf{Field Size of Algorithm~\ref{algo-1}:}

Algorithm~\ref{algo-1} requires the field size $|\Fq|>|\mA_r|$, and thus a base field $\Fq$ of order
\begin{align}\label{field_size_algo1}
q>\max\Big\{ \big|T\big|,~\big|\mA_{r}\big| \Big\}\footnotemark
\end{align}
\footnotetext{The reason for requiring $q > |T|$ here is to guarantee the existence of a $C_{\min}$-dimensional linear network code $\mC_{C_{\min}}$ on $G$.}is sufficient for constructing a family of local-encoding-preserving SLNCs with the fixed security level~$r$ and rates from $C_{\min}-r$ to $0$. In addition, we note that $\max\big\{|T|,~|\mA_{r}|\big\}$ is also the best known lower bound on the required field size for the existence of an SLNC with rate~$\w$ and security level~$r$ (cf.~\cite{GY-SNC-Reduction}). Therefore, we see that there is no penalty at all on the field size (in terms of the best known lower bound) for constructing such a family of local-encoding-preserving SLNCs.

\bigskip

\noindent\textbf{Complexity of Algorithm~\ref{algo-1}:}

For the purpose of determining the time complexity of Algorithm~\ref{algo-1}, we do not differentiate an addition from a multiplication over a finite field, although in general the time needed for a multiplication is much longer than that needed for an addition. We further assume that the time complexity of each operation, i.e., an addition or a multiplication, is $\mO(1)$ regardless of the finite field.

Now, we discuss the complexity of Algorithm~\ref{algo-1}.
\begin{itemize}
  \item For Line~1, in order to find $\w-1$ linearly independent vectors in $\vb_1^{\,(n-1)}$, $\vb_2^{\,(n-1)}$,$\cdots$, $\vb_{\w}^{\,(n-1)}$, it suffices to transform the $(n-1) \times \w$ matrix $\left[ \vb_1^{\,(n-1)} \ \ \vb_2^{\,(n-1)} \ \  \cdots \ \ \vb_\w^{\,(n-1)} \right]$ into one in row echelon form by a sequence of elementary row operations. By Gaussian elimination, this transformation takes at most $\mO\big( \w^2(n-1) \big)$ operations.
  \item For Line~2, in order to determine the linear relationship between the $n-1$ vectors $\vb_i^{\,(n-1)}$, $1\leq i \leq \w-1$,\footnote{Here, we still assume that $\vb_i^{\,(n-1)}$, $1\leq i \leq \w-1$, are chosen to be linearly independent.} $\vf_e^{\,(n-1)}$, $e\in A$, for each edge subset $A$ in $\mA_r$, it suffices to compute the rank of the $(n-1)\times(n-1)$ matrix $\left[ \vb_i^{\,(n-1)}, 1\leq i \leq \w-1,~\vf_e^{\,(n-1)}, e\in A \right]$, which takes at most $\mO\big( (n-1)^3 \big)$ operations by Gaussian elimination. By considering all the edge subsets $A$ in $\mA_r$, the complexity for partitioning $\mA_r$ into $\mA_r'$ and $\mA_r''$ (i.e., Line~2) is at most $\mO\big( |\mA_r|(n-1)^3 \big)$.
  \item For Line~3, we need to find a vector $\vh$ such that
      \begin{align*}
      \vh \in \Fq^{n-1}\setminus \bigcup_{A\in \mA_r'} \big(\mB_{\w-1}^{(n-1)}+\mL_A^{(n-1)}\big).
      \end{align*}
      By \cite[Lemma 11]{Yang-refined-Singleton}, such a vector can be found with
      $\mO\big( (n-1)|\mA_r'|\cdot\big[(n-1)^2+|\mA_r'|\big] \big)$ operations.
 \item We now consider the ``for'' loop (Lines 4--6). For Line~5, note that $\vb_i^{\,(n-1)}$, $1\leq i \leq \w-1$, $\vf_e^{\,(n-1)}$, $e\in A$, are $n-1$ linearly independent $(n-1)$-dimensional vectors for any $A\in \mA_r''$, and thus they together form a basis of the vector space $\Fq^{n-1}$. So, for the vector $\vh$ we have chosen, for each $A\in \mA_r''$, we can find the unique row $(n-1)$-vector $\big(\alpha_i, 1\leq i \leq \w-1,\ \beta_e, e\in A \big)\in \Fq^{n-1}$ such that
      \begin{align}\label{equ1-comp-algo-1}
      \vh=\sum_{i=1}^{\w-1}\alpha_i\vb_i^{\,(n-1)}+\sum_{e\in A} \beta_e\vf_e^{\,(n-1)}
      \end{align}
       with $\mO\big((n-1)^3\big)$ operations, since finding this row $(n-1)$-vector is equivalent to solving the system of linear equations \eqref{equ1-comp-algo-1}, which takes at most $\mO\big((n-1)^3\big)$ operations by Gaussian elimination. With the $(n-1)$-row vector $\big(\alpha_i, 1\leq i \leq \w-1,\ \beta_e, e\in A \big)\in \Fq^{n-1}$, we compute $\theta_A=\sum_{i=1}^{\w-1}\alpha_i b_{i,n}+\sum_{e\in A}\beta_e f_{e,n}$ in Line~6, which takes at most $\mO\big(n-1\big)$ operations. Thus, the total complexity of determining $\theta_A$ for all $A\in \mA_r''$ is at most $\mO\big( |\mA_r''|(n-1)^3 \big)$.

    \item For Line~7, to choose a nonzero element $\theta$ in $\Fq$ such that $\theta\cdot\theta_A\neq -1$ for all $A\in \mA_r''$, the complexity is $\mO\big( |\mA_r''|\big)$.

    \item For Line~8, the calculation of the vector $\vk=\theta\vh$ takes at most $\mO\big(n-1\big)$ operations.

  \item Clearly, the complexity of Lines~9~and~10 are at most $\mO\big( (n-1)|\Out(s)|\big)$ (that is $\mO (n-1)$ because $|\Out(s)|$ is fixed) and $\mO\big(n(n-1)\big)$, respectively.

  \item The complexity analysis of Line~11 is similar to that of Line~1. To obtain $Q^{(n-1)}$, it suffices to transform the $(n-1)\times n$ matrix $Q^{(n-1)}(\vk)$ into one in row echelon form, which takes at most $\mO\big((n-1)^2 \cdot n\big)$ operations by Gaussian elimination.
\end{itemize}
Therefore, by combining all the foregoing complexity analyses of Lines~1--11, the total complexity of Algorithm~\ref{algo-1} is at most
\begin{align}\label{complexity-algo1}
\mO\big( (n-1)^3 |\mA_r|+(n-1)|\mA_r'|^2 \big).
\end{align}

For the approach we proposed at the beginning of Section~\ref{subsec_secure-cond}, which, by totally redesigning a linear pre-coding operation at the source node, can also construct a local-encoding-preserving SLNC with the fixed security level $r$ and rate $\w-1$, the complexity is
\begin{align}\label{complexity-original-algo}
\mO\left( (\w-1) n^3\big|\mA_{r}\big|+(\w-1)(n-1)\big|\mA_{r}\big|^2+r(n-1)^3 \right)
\end{align}
(cf.~the~\ref{footnote}th footnote or Appendix~\ref{append-complx}).
Upon comparing \eqref{complexity-algo1} and \eqref{complexity-original-algo}, we see that the complexity of Algorithm~\ref{algo-1} is considerably smaller than one $\w$th of that of the approach described immediately above.
On the other hand, for Algorithm~\ref{algo-1}, in order to store the matrix $Q^{(n)}$ at the source node $s$, it suffices to store the column $(n-1)$-vector $\vk$ only. This implies that the storage cost is $\mO\big( n-1 \big)$, which is considerably smaller than the storage cost $\mO\big( n^2 \big)$ of the approach proposed at the beginning of Section~\ref{subsec_secure-cond} (cf.~the~\ref{footnote}th footnote). Thus, our approach reduces considerably the complexity and storage cost further.

%%%%%%%%%%%%%%%%%%%%%%%%%%%%%%%%%%%%%%%%%%%%%%%%%%%%%%%%%%%%%%%%%%%%%%%%%%%%

\section{Conclusions}\label{Sec_conclusion}

In a secure network coding system, the requirements for information transmission and information security may vary over time. We investigate the problem of secure network coding under the above consideration in this two-part paper. To efficiently solve this problem, we put forward local-encoding-preserving secure network coding, where a family of SLNCs is called local-encoding-preserving if all the SLNCs in this family share a common local encoding kernel at each intermediate node in the network. This approach can avoid all the shortcomings, in terms of computation cost, storage cost, and implementation overhead, of the straightforward but cumbersome approach that uses the existing code constructions to obtain an SLNC for each pair of rate and security level.

In this paper (i.e., Part~\Rmnum{1} of the two-part paper), we consider local-encoding-preserving secure network coding for a fixed security level.
We have developed an approach that can construct
a family of local-encoding-preserving SLNCs with a fixed security level and the rate ranging from 1 to the maximum possible.
We also presented a polynomial-time algorithm for efficient implementation.
Our approach not only guarantees the local-encoding-preserving property for the family of SLNCs as constructed, but also incurs no penalty on the required field size in the construction of such a family of local-encoding-preserving SLNCs for the existence of SLNCs in terms of the best known lower bound by Guang and Yeung~\cite{GY-SNC-Reduction}.

In Part~\Rmnum{2}~\cite{part2}, we will continue the studies in this paper by tackling first local-encoding-preserving secure network coding for a fixed rate,
and then local-encoding-preserving secure network coding for a fixed dimension (equal to the sum of rate and security level). The approaches in the current paper
and the companion paper will be combined to solve the ultimate problem of local-encoding-preserving secure network coding for the whole rate and security-level region.

%In addition, some interesting questions related to these problems remain open. For instance, when the network transmission suffers both wiretapping and errors possibly including random errors and adversary errors, how to deal with the similar problems and trade off information rate, security, and error-correcting capability. Perhaps some hints are available in \cite{Cai-StronglyGenericLNC-ISIT09, Ngai-Yeung-SNEC-NetCod09} and \cite{Guang-uni-MDS}. In addition, for the non-coherent wiretap networks, how to solve the corresponding problems and whether randomized network coding is feasible. Some works such as Ho \textit{et al.} \cite{Ho-Byzantine} and Jaggi \textit{et al.} \cite{Jaggi-Byzatine-IT08}, which considered randomized methods to defend against Byzantine attacks, may be helpful.

%%%%%%%%%%%%%%%%%%%%%%%%%%%%%%%%%%%%%%%%%%%%%%%%%%%%%%%%%%%%%%%%%%%%%%%%%%%%%%%%%%%%%%%%%%%%%
\numberwithin{thm}{section}
\appendices

\section{Complexity Analysis of Constructing a Required Matrix $Q^{(n)}$}\label{append-complx}

Let $\w$ be the fixed rate and $r$ be the security level. Let $n=\w+r\leq C_{\min}$. For a given $n$-dimensional linear network code $\mC_{n}=\big\{\vf_e^{\,(n)}:~e\in E \big\}$ over a finite field $\Fq$  on the network $G$,\footnote{Here, we assume the size of $\Fq$ sufficiently large to guarantee the existence of an $n$-dimensional linear network code and an $n$-dimensional SLNC with rate $\w$ and security level $r$ on the network $G$.} we consider in the following the complexity of constructing an $\Fq$-valued $n\times n$ invertible matrix $Q^{(n)}=\Big[ \vb_1^{\,(n)}\ \ \vb_2^{\,(n)}\ \ \cdots \ \ \vb_{n}^{\,(n)} \Big]$ using the method in \cite{Cai-Yeung-SNC-IT} such that $(Q^{(n)})^{-1}\cdot\mC_{n}$ is a rate-$\w$ and security-level-$r$ SLNC on $G$, i.e., \rmnum{1}) $\vb_1^{\,(n)}$, $\vb_2^{\,(n)}$, $\cdots$, $\vb_{n}^{\,(n)}$ are linearly independent and \rmnum{2})
\begin{align}\label{append-complx_equ1}
\big\langle  \vb_i^{\,(n)}:\ 1\leq i \leq \w  \big\rangle \bigcap
\big\langle \vf^{\,(n)}_e:\ e\in A \big\rangle=\{\bzero\},
\quad \forall~A\in \mA_r.
\end{align}

In order to constructing a matrix $Q^{(n)}$, we in turn choose $n$ vectors $\vb^{\,(n)}_i$, $1\leq i \leq n$, as follows:
\begin{align}
& \vb^{\,(n)}_1 \in \Fq^n \setminus \bigcup_{A\in \mA_{r}} \mL_A^{(n)}, \label{append-complx_vb_1}\\
& \vb^{\,(n)}_i \in \Fq^n \setminus \bigcup_{A \in \mA_{r}} \big(\mB_{i-1}^{(n)}+\mL_A^{(n)}\big),\quad 2\leq i \leq \w, \label{append-complx_vb_i}\\
& \vb^{\,(n)}_i \in \Fq^n \setminus \mB_{i-1}^{(n)}, \quad \w+1\leq i \leq n,\label{append-complx_vb_j}
\end{align}
where $\mB_i^{(n)}=\big\langle  \vb_j^{\,(n)}:~1\leq j \leq i \big\rangle$ and $\mL_A^{(n)}=\big\langle  \vf_e^{\,(n)}:~e\in A \big\rangle$. Clearly, the chosen $\vb^{\,(n)}_i$, $1\leq i \leq n$, are linearly independent and satisfy the security condition \eqref{append-complx_equ1} for security level $r$.

By \cite[Lemma 11]{Yang-refined-Singleton}, the vectors $\vb^{\,(n)}_i$, $1\leq i \leq \w$ satisfying \eqref{append-complx_vb_1} or \eqref{append-complx_vb_i} can be found in time
      \begin{align*}
      \mO\big( \w n^3\big|\mA_{r}\big|+\w n\big|\mA_{r}\big|^2 \big),
      \end{align*}
and the vectors $\vb^{\,(n)}_i$, $\w+1 \leq i \leq n$ satisfying \eqref{append-complx_vb_j} can be found in $\mO\big( r(n^3+n) \big)$ operations.
Therefore, the total complexity is
\begin{align*}
\mO\left( \w n^3\big|\mA_{r}\big|+\w n\big|\mA_{r}\big|^2+rn^3 \right).
\end{align*}

%%%%%%%%%%%%%%%%%%%%%%%%%%%%%%%%%%%%%%%%%%%%%%%%%%%%%%%%%%%%%%%%%%%%%%%%%%%%

\end{document}